\documentclass[12pt]{article}
\usepackage{amsmath,amssymb,epsfig}
\usepackage{refcheck}
\makeatletter \@addtoreset{equation}{section} \makeatother
\renewcommand{\theequation}{\thesection.\arabic{equation}}
\newcommand{\ba}{\begin{array}}
\newcommand{\ea}{\end{array}}
\newcommand{\beq}{\begin{equation}}
\newcommand{\eeq}{\end{equation}}
\newcommand{\bea}{\begin{eqnarray}}
\newcommand{\eea}{\end{eqnarray}}

\def\bce{\begin{center}}
\def\ece{\end{center}}

\def\nonu{\nonumber}

\def\pa{\partial}
\def\al{\alpha}
\def\be{\beta}
\def\ga{\gamma}

\def\de{\delta}

\def\ep{\epsilon}

\def\la{\lambda}

\def\si{\sigma}

\def\diag{\mathop{\rm diag}}

\def\eps6{{\displaystyle \mathop{\epsilon}^{6}}{}}

\def\nab6{{\displaystyle \mathop{\nabla}^{6}}{}}

\def\0{{\sst{(0)}}}
\def\1{{\sst{(1)}}}
\def\2{{\sst{(2)}}}
\def\3{{\sst{(3)}}}
\def\4{{\sst{(4)}}}
\def\5{{\sst{(5)}}}
\def\6{{\sst{(6)}}}
\def\7{{\sst{(7)}}}
\def\8{{\sst{(8)}}}

\def\ba{\begin{array}}
\def\ea{\end{array}}
\def\beq{\begin{equation}}
\def\eeq{\end{equation}}
\def\be{\begin{equation}}
\def\ee{\end{equation}}

\def\Tr{\mathop{\rm Tr}}

\def\diag{\mathop{\rm diag}}

\def\la{\lambda}
\def\eps{\epsilon}

\def\ba{\begin{array}}
\def\ea{\end{array}}
\def\beq{\begin{equation}}
\def\eeq{\end{equation}}
\def\be{\begin{equation}}
\def\ee{\end{equation}}

\def\Tr{\mathop{\rm Tr}}

\def\diag{\mathop{\rm diag}}

\def\la{\lambda}
\def\eps{\epsilon}

\newcommand{\bean}{\begin{eqnarray*}}
\newcommand{\eean}{\end{eqnarray*}}

\begin{document}
\thispagestyle{empty} \addtocounter{page}{-1}
   \begin{flushright}
\end{flushright}

\vspace*{1.3cm}
  
 \centerline{ \Large \bf  ${\cal N}=8$ Gauged Supergravity  Theory and } 
\vspace{.3cm} 
\centerline{ \Large \bf  
${\cal N}=6$ Superconformal Chern-Simons Matter Theory   } 
\vspace*{1.5cm}
\centerline{\bf Changhyun Ahn {\rm and} Kyungsung Woo }
\vspace*{1.0cm} 
\centerline{\it  
Department of Physics, Kyungpook National University, Taegu
702-701, Korea} 
\vspace*{0.8cm} 
\centerline{\tt ahn@knu.ac.kr
} 
\vskip2cm

\centerline{\bf Abstract}
\vspace*{0.5cm}

By studying the previously known holographic ${\cal N}=4$
supersymmetric renormalization group flow(Gowdigere-Warner) in four dimensions,
we find the mass deformed Chern-Simons matter theory which has ${\cal
N}=4$ supersymmetry 
by adding the four mass terms among eight adjoint fields.    
The geometric superpotential from the eleven dimensions is found and 
provides the M2-brane probe analysis.
As second example, we consider known holographic ${\cal N}=8$
supersymmetric renormalization group flow(Pope-Warner) in four dimensions.
The eight mass terms are added and similar geometric superpotential is 
obtained.

\baselineskip=18pt
\newpage
\renewcommand{\theequation}
{\arabic{section}\mbox{.}\arabic{equation}}

\section{Introduction}

The
three dimensional ${\cal N}=6$ Chern-Simons matter theories
with gauge group $U(N) \times U(N)$ and level $k$ 
have been studied in \cite{ABJM}.
This theory is described as the low energy limit of $N$ M2-branes at 
${\bf C}^4/{\bf Z}_k$ singularity.
Since the coupling of this theory may be thought of as $\frac{1}{k}$
and so this is weakly coupled for large $k$. 
In particular, for $k=1, 2$, the full ${\cal N}=8$ supersymmetry( 
$SO(8)$ R-symmetry) is preserved and the theory is strongly coupled. 

The renormalization group(RG) flow
between the ultraviolet(UV) fixed point and 
the infrared(IR) fixed point of the three dimensional 
field theory has a close connection with a gauged supergravity solution in
four dimensions. 
There exist holographic
RG flow equations connecting ${\cal N}=8$ $SO(8)$ fixed point 
to ${\cal N}=2$ $SU(3) \times U(1)$ fixed point \cite{AP,AW}(See also \cite{NW}).
Moreover,  
the other holographic
RG flow equations from ${\cal N}=8$ $SO(8)$ fixed point 
to
${\cal N}=1$ $G_2$
fixed point also exist \cite{AW,AI,AR99}(See also \cite{Warner83,Warner84,AI02}).
The exact solutions to the $M$-theory lift of these
RG flows are known in \cite{CPW,AI}.

The mass deformed $U(2) \times U(2)$
Chern-Simons matter theory with level $k=1$ 
or $k=2$ preserving ${\cal N}=2$ $SU(3) \times U(1)$ symmetry was studied 
in \cite{Ahn0806n2,BKKS,KKM}. For ${\cal N}=1$ $G_2$ symmetric case, 
the corresponding mass deformation was described in \cite{Ahn0806n1}.
Recently \cite{Ahn0812} the nontrivial nonsupersymmetric 
flow equations preserving $SO(7)^{\pm}$ have been described in the
context of ${\cal N}=6$ superconformal Chern-Simons matter theory. 
In \cite{Ahn0812}, the possibility having the flow equations from 
${\cal N}=1$ $G_2$ fixed point to ${\cal N}=2$ $SU(3) \times U(1)$
fixed point, by realizing 1) the negativity of mass term around $G_2$
fixed point and 2) 
the natural symmetry breaking $G_2 \rightarrow SU(3)$, was 
proposed. 
Very recently, this possibility was shown in \cite{BHPW} by
introducing a new superpotential. 

As mentioned and suggested in \cite{Ahn0812}, it is known that  
there exist ${\cal N}=4$ supersymmetric flows with equal mass terms
for four real scalars in gauged supergravity theory discovered by
Gowdigere and Warner \cite{GW}.  
All the previous  flow equations were obtained from  
$SU(3)$-invariant
sectors
of gauged ${\cal N}=8$ supergravity in four dimensions.
Due to the $SU(2) \times SU(2)$ R-symmetry of ${\cal N}=4$ 
supersymmetry, this ${\cal N}=4$ supersymmetric flows are not
contained in $SU(3)$-invariant sectors.  
We would like to see what is the gauge dual to ${\cal N}=4$
supersymmetric flows in four dimensions \cite{GW}.
Moreover, there exist  ${\cal N}=8$ 
supersymmetric flows with equal mass terms
for eight real scalars discovered by
Pope and Warner \cite{PW}. In this case, the R-symmetry is given by 
$SO(4) \times SO(4)$. We also find out the gauge dual to 
${\cal N}=8$
supersymmetric flows in four dimensions \cite{PW}.

In this paper, 
starting from the first order differential equations, that are the
${\cal N}=4$ 
supersymmetric flow solutions in four dimensional ${\cal N}=8$ gauged
supergravity
interpolating between an exterior $AdS_4$ region with maximal
${\cal N}=8$ supersymmetry
and an interior region with ${\cal N}=4$ supersymmetry,  
we would like to interpret this as the RG flow in Chern-Simons matter theory
broken to the mass-deformed Chern-Simons matter theory 
by the addition of mass terms for the adjoint superfields. 
An exact correspondence 
may be obtained between fields of bulk supergravity in the $AdS_4$ region
in four-dimensions 
and composite operators of the IR field theory in three-dimensions. 
The three dimensional analog of Leigh-Strassler \cite{LS}
RG flow in the context of mass-deformed Chern-Simons matter 
theory is expected  but the present RG flows do not reveal the marginality
of Leigh-Strassler because there are no fixed critical points.   
As analyzed in \cite{FGPW} where they observed the work of  \cite{LS},
the beta function description in the context of 3-dimensional
Chern-Simons matter theory can be done.   
We present the results of probing the eleven-dimensional supergravity
solution corresponding to RG flows.

In section 2, we review the supergravity solution in four dimensions
in the context of RG flow and describe the scalar potential and the superpotential. 
The analysis for the supersymmetry variations on 
the spin $\frac{1}{2}$ and $\frac{3}{2}$ fields is new. 
In section 3, we deform BL theory \cite{BL0711,BL06,BL0712} 
by adding four mass terms and write down
the $[SU(2) \times U(1)]^2$-invariant 
superpotential in ${\cal N}=2$ superfields.
We also describe the corresponding mass-deformed Chern-Simons matter theory.  
In section 4, 
the 11-dimensional geometric 
superpotential which reduces to the usual $AdS_4$ superpotential for the
particular internal coordinate is computed and this is 
needed to analyze the M2-brane probe analysis.

In section 5, we review the supergravity solution in four dimensions
in the context of RG flow.
The analysis for the supersymmetry variations on 
the spin $\frac{1}{2}$ and $\frac{3}{2}$ fields is made. 
In section 6, we deform BL theory and Chern-Simons matter theory 
by adding eight mass terms and write down
the $SO(4) \times SO(4)$-invariant 
superpotential in ${\cal N}=2$ superfields.  
In section 7, 
the 11-dimensional geometric 
superpotential  is described for the M2-brane 
probe analysis.

In section 8, we present the future directions.
In the appendices, we present the details which are needed for the
previous sections \footnote{There are other related works 
\cite{diffsusy,CDS,BILPSZ,Ahn0810,ADFGT,JT} which 
discuss about the different supersymmetric theories in the context of
Chern-Simons matter theory.}. 

\section{The holographic ${\cal N}=4$ RG flow 
in four dimensions}

The invariant scalar manifold by Gowdigere-Warner \cite{GW} 
consisting of eight scalars 
has a factor $SO(6,1) \times SL(2,{\bf R})$ in $E_{7(7)}$ which has a
maximal non-compact subgroup $SL(8,{\bf R})$.
The special gauge choice reduces this scalar manifold to 
$SL(2, {\bf R}) \times SL(2, {\bf R})$ and eight scalars go to four.
Each $SL(2,{\bf R})$ has two supergravity scalars. 
The noncompact generators of $SL(2, {\bf R}) \times SL(2, {\bf R})$ 
can be written as
\cite{GW}
\bea
\phi_{ijkl} & = & \frac{1}{2} (\alpha \cos \phi + i \alpha \sin \phi)
\delta_{ijkl}^{1234} +\frac{1}{2}(\alpha \cos \phi - i \alpha \sin \phi)
\delta_{ijkl}^{5678} 
\nonu \\
&+ & \frac{1}{2}
(\chi \cos \varphi + i \chi \sin \varphi) 
(\delta_{ijkl}^{1256} + \delta_{ijkl}^{3456}) 
+\frac{1}{2} (\chi \cos \varphi - i \chi \sin \varphi) 
(\delta_{ijkl}^{1278} + \delta_{ijkl}^{3478}), 
\label{phiijkl}
\eea
where $\alpha e^{i \phi}$ parametrizes the one $SL(2,{\bf R})$ and 
$\chi e^{i \varphi}$ parametrizes the other $SL(2,{\bf R})$.
The real parts of these correspond to the ${\bf 35}_v$ of $SO(8)$
and the imaginary parts of these correspond to the ${\bf 35}_c$ of $SO(8)$.
Then 56-beins ${\cal V}(x)$ can be written as $ 56\times 56$ matrix of
${\cal N}=8$ supergravity in four-dimensions  whose
elements are some functions of scalar, pseudo-scalars, $\phi$ and $\varphi$
out of seventy fields \cite{CJ}
by exponentiating
the vacuum expectation value $\phi_{ijkl}$ through
\bea
{\cal V}(x)=
\mbox{exp} \left(
\begin{array}{cc}
0 &  2 \phi_{ijkl}(x)  \\
 2 \phi^{ijkl}(x) & 0 
\end{array} \right).
\label{calV}
\eea   
On the other hand, 28-beins $u$
and $v$ of ${\cal N}=8$ supergravity are 
elements of this ${\cal V}(x)$ according to 
\bea
{\cal V}(x)=
\left(
\begin{array}{cc}
u_{ij}^{\;\;IJ}(x) & v_{ijKL}(x)  \\
v^{klIJ}(x) & u^{kl}_{\;\;KL}(x)
\end{array} 
\right). 
\label{56bein}
\eea
One can construct these 28-beins $u$ and $v$ 
in terms of  $\alpha, \chi, \phi$ and 
$\varphi$ using (\ref{phiijkl}), (\ref{calV}) and (\ref{56bein})
explicitly and they are given in (\ref{uv}) of the appendix A. 
Now it is ready to get the complete expression for $A_1$ and 
$A_{2}$ tensors in terms of $\alpha, \chi, \phi$ and 
$\varphi$ from T-tensor \cite{dN82,dN82-1}. 

It turns out that 
$A_1$ tensor has three distinct complex
eigenvalues, $z_1$, $z_2$ and $z_3$ 
with degeneracies 4, 2, and 2 respectively and has the following form \cite{GW}
\begin{equation}
A_1 =\mbox{diag} \left(z_1, z_1, z_1, z_1, z_2,
z_2, z_3, z_3 \right), 
\nonu
\end{equation}
where the eigenvalues are functions of $\al, \chi, \phi$ and $\varphi$:
\bea
z_1 & = & \cosh \al \; \cosh^2 \chi + e^{-i \phi} \sinh \al \; \sinh^2 \chi,
\nonu \\
z_2 & = &  \cosh \al \; \cosh^2 \chi + e^{-i(2\varphi- \phi)} \sinh
\al \; \sinh^2 \chi,
\nonu \\
z_3 & = &  \cosh \al \; \cosh^2 \chi + e^{i(2\varphi+ \phi)} \sinh \al
\; \sinh^2 \chi.
\label{zis}
\eea
One of the eigenvalues of
$A_1$ tensor, $z_1$,  will provide a ``superpotential'' of scalar
potential $V$ and
be crucial for the analysis of domain-wall solutions.   
Of course, the choice for the other $z_2$ or $z_3$ as a superpotential
will give rise to other theory which has different supersymmetry.
We are mainly concerned with the choice of $z_1$ in this paper. 

Similarly, $A_{2}$ 
tensor can be obtained from the T-tensor(triple product of
$u$ or $v$ fields: the explicit form is given by (\ref{a1a2})). 
It turns out that they are written as seven-kinds of 
fields
$y_i$ where $i=1, \cdots, 7$ and are given in (\ref{A2}) and 
(\ref{y7}) of the appendix A
where some of these are related to 
the derivatives of eigenvalues of $A_1$ tensor $z_1, z_2$ or $z_3$ 
with respect to
$\al$ and $\chi$.

Finally, 
the scalar potential \cite{GW} from the general expression of \cite{dN82}
can be written, by adding all the components of
$A_1, A_2$ tensors,  as
\bea
V(\al, \chi) & = & 
-g^2 \left( \frac{3}{4} \left| A_1^{\;ij} \right|^2-\frac{1}{24} \left|
A^{\;\;i}_{2\;\;jkl}\right|^2 \right)=
- 2g^2 \left[ \cosh (2\alpha) +2
\cosh (2 \chi) \right]  \nonu \\
&=& g^2 \left( 2 \left|\frac{\partial
W}{\partial \al} \right|^2 +  \left|\frac{\partial W}{\partial
\chi} \right|^2 - 6  \left|W \right|^2 \right),
\label{pot}
\eea
where we consider the particular case
\bea
\phi=0, \qquad \varphi =\frac{\pi}{2},
\label{cond}
\eea
and the superpotential \cite{GW} is given by
\bea
W(\al, \chi) = \frac{1}{2} \left[ \frac{1}{\rho} 
+ \rho \cosh (2\chi) \right]=z_1, \qquad
\rho \equiv e^{\alpha}.
\label{superpot}
\eea
The other four components for $A_1$ are given by
$\frac{1}{2}\left( \rho - \rho^{-1} \cosh 2\chi \right)=z_2=z_3$. 
This implies half-maximal supersymmetry(that is, 
${\cal N}=4$ supersymmetry: this will be clear when we understand the
variations of spin $\frac{1}{2}$ and spin $\frac{3}{2}$ fields later).
The scalar potential at the $SO(8)$ UV critical point where $\alpha=0$ 
and $\chi=0$ becomes 
$V=-6g^2$ where $g=\frac{1}{\sqrt{2} L}$ and the superpotential 
$W$ becomes 1. 
Note that the supergravity scalar potential (\ref{pot}) is independent of the
angles $\phi$ and $\varphi$ even without the conditions (\ref{cond}).
This critical point is common to both a scalar 
potential which has a maximum value
and a superpotential (\ref{superpot}) 
which has a minimum value and is depicted in Figure 1.
The choice $\phi=0=\varphi$ where all the $z_i$'s in (\ref{zis}) are equal 
preserves ${\cal N}=8$ supersymmetry.

\begin{figure}
\begin{center}
\begin{minipage}[t]{6.8cm}
\centerline{\hbox{\psfig{file=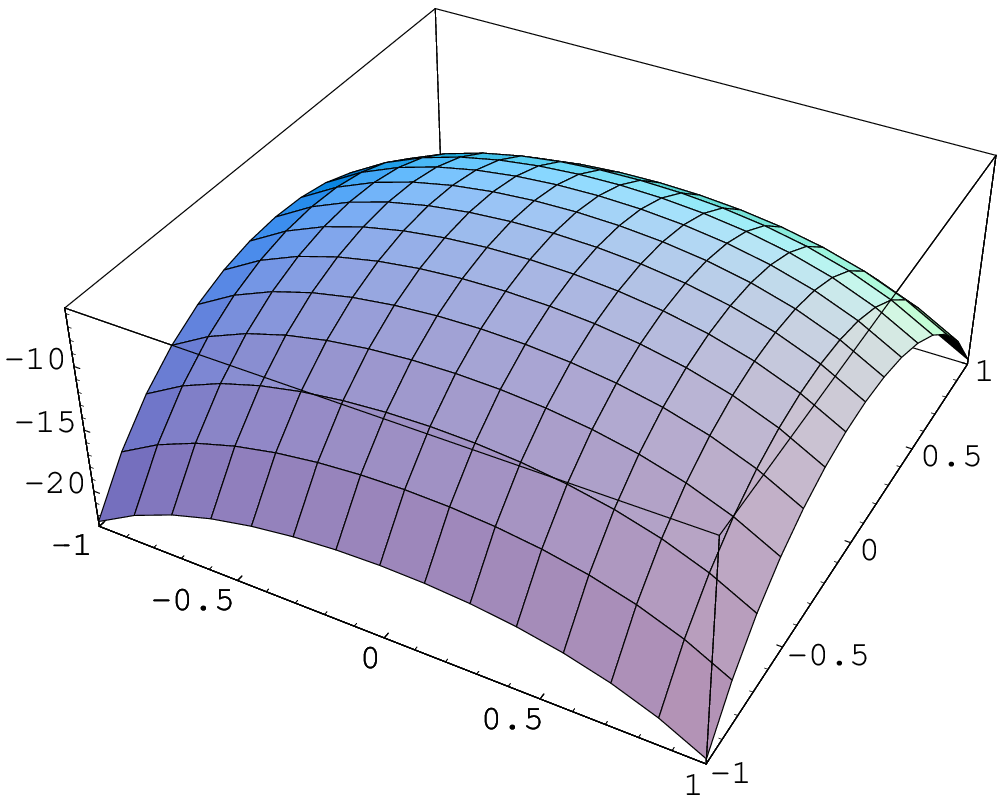,width=6.8cm}}}
\end{minipage}\hspace{1cm}
\begin{minipage}[t]{6.8cm}
\centerline{\hbox{\psfig{file=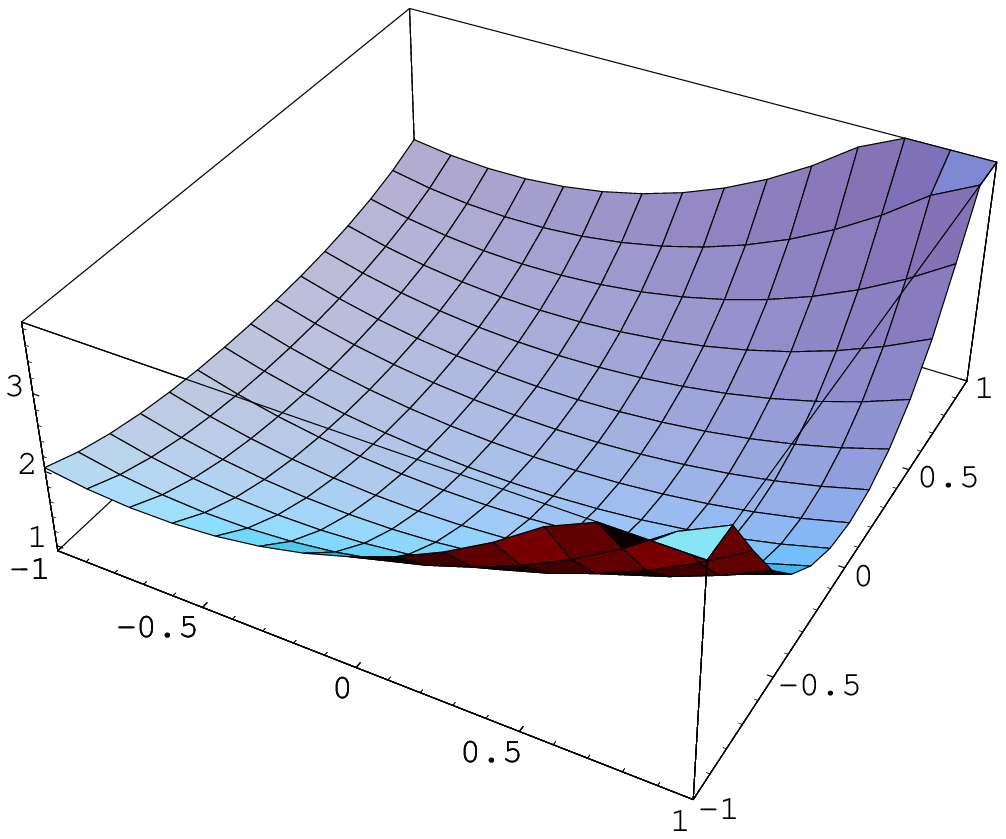,width=6.8cm}}}
\end{minipage}
\caption{\sl Three dimensional plots of 
the scalar potential $V=-2g^2 \left( \cosh 2\alpha +2
\cosh 2 \chi \right) $(left)  given by (\ref{pot}) and 
the superpotential $W=\frac{1}{2} 
\left( e^{-\alpha} + e^{\alpha} \cosh 2\chi \right)$(right) 
given by (\ref{superpot}) in the 4-dimensional gauged supergravity. 
The axes $(\alpha, \chi)$ are two vevs that 
parametrize the $[SU(2) \times U(1)]^2$ 
invariant manifold in the 28-beins of the theory. 
The $SO(8)$ UV critical point is located at $\alpha=\chi=0$ and has
$V=-6$ and $W=1$.
We have set the $SO(8)$ gauge coupling $g$ in the scalar potential as 
$g=1$.   }
\end{center}
\end{figure}

The resulting Lagrangian of the 
scalar-gravity sector can be obtained 
by finding out the scalar kinetic terms
appearing in the action in terms of $\al, \chi, \phi$ and 
$\varphi$. 
By taking the product of $  A_{\mu}$ given in (\ref{amu}) of appendix
A and
its complex conjugation and taking into account the 
multiplicity four,  
we arrive at the following scalar kinetic term \cite{GW} with (\ref{cond})
\bea
(\pa_\mu \alpha)^2 + 2(\pa_\mu \chi)^2.
\label{kineticterm}
\eea
By substituting the domain-wall ansatz 
\bea
ds^2= e^{2A} \eta_{\mu \nu} dx^{\mu} dx^{\nu} + dr^2
\nonu
\eea
with $
\eta_{\mu \nu}=(-,+,+)$
into the Lagrangian, 
the Euler-Lagrangian equations are given in terms of 
the functional $E[A, \al, \chi]$ \cite{ST}.
Then the 
energy-density per unit area transverse to $r$-direction is given by
\bea
 E[A,\alpha,\chi]  =  \int_{-\infty}^{\infty} dr e^{3A}
\left[- 3 \left( 2 (\pa_r A )^2 + \pa_r^2 A \right) - 
\left( \partial_r \al \right)^2 -2 \left( \partial_r \chi \right)^2
  -V(\al, \chi)\right].
\nonu
\eea
Then $E[\al, \chi]$ is extremized by the 
BPS domain-wall solutions. The first order differential
equations are the gradient flow equations
of a superpotential defined on a restricted two-dimensional slice of
the scalar manifold and simply related to the potential of gauged supergravity
on this slice via (\ref{pot}) and (\ref{superpot}).
The half-maximal(${\cal N}=4$) supersymmetric flow satisfying the 
first order differential equations for given superpotential $W$ 
in (\ref{superpot}) 
leads to the solutions \cite{GW} for $\alpha(r), \chi(r), A(r)$:
\bea  
\frac{d \al}{d r}  =  -
\sqrt{2} \, g \, \partial_{\al} W ,\qquad
\frac{d \chi}{d r}   = -  
\frac{\sqrt{2}}{2} \, g \, \pa_{\chi} W, \qquad
\frac{d A}{d r}  =   \sqrt{2} \, g \, W.
\label{flow}
\eea
The
flow equations imply that 
\bea
\alpha = \frac{1}{2} \log \left[
  e^{-2\chi}+
\gamma \sinh (2\chi) \right] \qquad \mbox{and} \qquad
 e^{A} = 
\frac{k \rho}{\sinh(2\chi)}, 
\label{equa1}
\eea
where $\gamma$ and $k$ are
constants of the integration \cite{GW}.  
In particular, for $\gamma =0$, $\alpha=-\chi$ from the first equation
of (\ref{equa1}) all along the flow and $e^A$ from the second equation
of (\ref{equa1}) goes to zero as $\chi
\rightarrow \infty$. Note that the superpotential (\ref{superpot})
is invariant under $\chi \rightarrow -\chi$. There exists similar
behavior in \cite{PW1} from ten-dimensional type IIB string theory.

One can understand the BPS bound, so-called inequality of 
the energy-density as a consequence of supersymmetry preserving 
bosonic background. 
For the supersymmetric bosonic backgrounds, the variations
of spin-$\frac{1}{2}$ and spin-$\frac{3}{2}$- fields should vanish. 
The gravitational and scalar parts of these variations are \cite{dN82}:
\bea
\de\psi_{\mu}^i & = & 
 2 D_{\mu} \epsilon^i - \sqrt{2} g A_1^{\;ij} \ga_{\mu} \ep_j, 
\nonu \\
\de \chi^{ijk}  & = & - \ga^{\mu} A_{\mu}^{\;\;ijkl} \ep_l- 2 g 
A_{2l}^{\;\;\;ijk} 
\ep^l,
\label{susy}
\eea
where the covariant derivative on supersymmetry parameter is given by
\bea
 D_{\mu} \epsilon^i = \partial_{\mu} \ep^i - \frac{1}{2} 
\omega_{\mu a b} \sigma^{ab} \ep^i
+ \frac{1}{2} 
{\cal B}_{\mu\;\;\;j}^{\;\;i} \ep^j, \;\;\; 
{\cal B}_{\mu\;\;\;j}^{\;\;i} \equiv \frac{2}{3} \left( u^{ik}_{\;\;\;IJ}
\partial_{\mu} u_{jk}^{\;\;\;IJ} - v^{ikIJ} \partial_{\mu} v_{jkIJ} \right).
\label{cov}
\eea
Here $\ep_i$ and $\ep^j$ are complex conjugates each other under the chiral 
basis. 
In this basis, the $\gamma$ matrices satisfy $\{\gamma^{\mu}, 
\gamma^{\nu} \} = 2 g^{\mu \nu}$ with $g_{\mu \nu}
=\mbox{diag}(-e^{2A}, e^{2A}, e^{2A}, 1)$ and  $\gamma^i =
\left(
\begin{array}{cc}
0 & i \sigma^i \nonu \\
-i \sigma^i & 0 \\
\end{array}
\right)$ where $\sigma^i$ are Pauli matrices and
$\gamma^5 = i \gamma^0 \gamma^1 \gamma^2 \gamma^3$. Also 
there is $\gamma^0 =
\left(
\begin{array}{cc}
0 & -i {\bf 1} \nonu \\
-i {\bf 1} & 0 \\
\end{array}
\right)$.
The field ${\cal B}_{\mu\;\;\;j}^{\;\;i}$ is a $SU(8)$ gauge field for a local
$SU(8)$ invariance, $\omega$ a spin connection, $\sigma$ a commutator
of two $\gamma$ matrices $\sigma^{ab} =\frac{1}{4} [\gamma^a, \gamma^b]$. 
Under the projection operators $(1\pm \gamma_5)/2$
the supersymmetry parameter 
$\ep^i$ has the four column components as $(\eta_1, \eta_2,
0, 0)$ where $\eta_1, \eta_2$ are complex spinor fields.
Moreover, complex conjugate 
$\ep_i$ is the charge conjugate spinor of $\ep^i$ and satisfies
$\ep_i = C {\gamma^0}^{T} \ep^{i \ast}$ 
and has the 4 column components as $(0, 0, \eta_3=i {\eta_2}^{\ast}, 
\eta_4=-i{\eta_1}^{\ast})$ with $C =
\left(
\begin{array}{cc}
i \sigma^2 & 0 \nonu \\
0 & -i\sigma^2 \\
\end{array}
\right)$.
The explicit form for ${\cal B}_{\mu\;\;\;j}^{\;\;i}$ is presented in
appendix A (\ref{calB}).

The variation of 56 Majorana spinors $\chi^{ijk}$ gives rise to
the first order differential equation of $\al$ and 
$\chi$ by exploiting
the explicit forms of $ A_{\mu}^{\;\;ijkl}$ (\ref{amu}) and 
$A_{2l}^{\;\;\;ijk}$ (\ref{A2}) and (\ref{y7}) in the
appendix A.
Although there is a summation over the last index $l$ appearing in
$ A_{\mu}^{\;\;ijkl}$ and $A_{2l}^{\;\;\;ijk}$ in the right hand side of
(\ref{susy}), this structure implies
that summation runs over only one index. 
The vanishing of variation of  $\chi^{ijk}$ for supersymmetry
parameter $\epsilon^i$ where $i=1,2,3,4$ 
leads to 
\bea
\sqrt{2} e^{-i\phi} \left[\partial_{\mu}\alpha -
\frac{i}{2}\sinh\left(2\alpha\right)
\partial_{\mu}\phi \right] \left(
\begin{array}{c}
\eta_2^{\ast}  \\
\eta_1^{\ast}  \\ 
\end{array}
\right)
& = & -2 g e^{i\phi} \frac{\partial z_{1}^{*}}{\partial\alpha}
\left(\begin{array}{c}
\eta_1  \\
\eta_2  \\
\end{array}
\right),
\nonu \\ 
\sqrt{2}  e^{-i\varphi} \left[ \partial_{\mu}\chi -\frac{i}{2}
\sinh\left(2\chi\right) \partial_{\mu}\varphi \right] 
\left(
\begin{array}{c}
\eta_2^{\ast}  \\
\eta_1^{\ast}  \\
\end{array}
\right)
& = &
-g  e^{-i\varphi} \frac{\partial z_{1}^{*}}{\partial\chi}
\left( \begin{array}{c}
\eta_1  \\
\eta_2  \\
\end{array}
\right),  \nonu \\
\sqrt{2}  e^{i\varphi} \left[ \partial_{\mu}\chi + \frac{i}{2}
\sinh\left(2\chi\right) \partial_{\mu}\varphi \right] 
\left(
\begin{array}{c}
\eta_2^{\ast} \\
\eta_1^{\ast}  \\
\end{array}
\right) 
& = &
-g e^{i\varphi} \frac{\partial z_{1}^{*}}{\partial\chi}
\left(\begin{array}{c}
\eta_1  \\
\eta_2  \\
\end{array}
\right).
\label{first} 
\eea

Moreover, the variation of gravitinos 
$\psi_{\mu=0,1,2}^i$ 
will lead to
\bea
\partial_{r}A \left(\begin{array}{c}
\eta_1  \\
\eta_2 \\
\end{array}
\right)  =  \sqrt{2}g z_{1} \left(
\begin{array}{c}
\eta_2^{\ast}  \\
\eta_1^{\ast} \\
\end{array}
\right).
\label{second}
\eea
Finally, one of the variation of gravitinos $\psi_{\mu=3}^i$ gives rise to
\bea
2\partial_{r}  \left(\begin{array}{c}
\eta_1 \\
\eta_2 \\
\end{array}
\right) = i \partial_{\mu}\phi
\sinh^{2} \alpha  \left(\begin{array}{c}
\eta_1 \\
\eta_2 \\
\end{array}
\right)
-\sqrt{2}g z_{1}  \left(
\begin{array}{c}
\eta_2^{\ast}  \\
\eta_1^{\ast} \\
\end{array}
\right).
\label{third}
\eea

Let us denote the complex spinor fields in terms of their magnitudes
and phases as follows: $\eta_1 =|\eta_1(r)| e^{i \beta(r)}$ and
$\eta_2=
|\eta_2(r)| e^{i \delta(r)}$. By multiplying the $\eta_2$ into the
first row and $\eta_1$ into the second row in the first equation 
(\ref{first}) and
combining these two relations, one obtains
$|\eta_1(r)|=|\eta_2(r)|\equiv |\eta(r)|$.
Let us consider the equation (\ref{third}) and substitute the $\eta_1$
and $\eta_2$. By multiplying $e^{-i\beta}$ into the first row and 
$e^{-i\delta}$ into the second row and subtracting these two, then 
one obtains $(\pa_r \beta -\pa_r \delta)\eta$ which implies that 
$\beta = \delta + \mbox{const}$.   
Now let us analyze the equation (\ref{second}). From the first row, 
one gets $\pa_r A = \sqrt{2} g z_1 e^{-2i\beta}$. This implies that 
$z_1 e^{-2i\beta}$ should be real because $\pa_r A$ is real.

Let us compare the second and third equations of (\ref{first}).
One gets 
\bea
\sqrt{2}  e^{-i\varphi} \left[ \partial_{\mu}\chi - \frac{i}{2}
\sinh\left(2\chi\right) \partial_{\mu}\varphi \right] & = & -g  
e^{-i\varphi+2i\beta} \frac{\partial z_{1}^{*}}{\partial\chi},
\nonu \\
\sqrt{2}  e^{i\varphi} \left[ \partial_{\mu}\chi + \frac{i}{2}
\sinh\left(2\chi\right) \partial_{\mu}\varphi \right] & = & -g  
e^{i\varphi+2i\beta} \frac{\partial z_{1}^{*}}{\partial\chi}.
\label{fourth}
\eea
Then we obtain $e^{2i\beta} \frac{\partial 
z_{1}^{*}}{\partial\chi} = e^{-2i\beta} 
\frac{\partial z_{1}}{\partial\chi}$ because the left hand side 
of second equation of (\ref{fourth}) is a complex conjugation of the
left hand side of first equation. This leads to the fact that 
$e^{2i\beta} \frac{\partial 
z_{1}^{*}}{\partial\chi}$ is real. Therefore the expression of the
bracket in (\ref{fourth}) should be real. So one has $\pa_r \varphi =0$.
Moreover by plugging the expression of $z_1$, one has 
$\cosh \alpha \sin (2\beta) + \sinh \alpha \sin (2\beta +\phi) =0$.
One simple solution for this gives rise to $\beta=0$ and $\phi=0$.

By collecting all the informations so far, one obtains 
$\pa_r A = \sqrt{2} g W$ because $z_1$ becomes $W$. From
(\ref{first}),
one also obtains $\pa_r \alpha = -\sqrt{2} g \pa_{\alpha} W$ and 
$\pa_r \chi =-\frac{1}{\sqrt{2}} g \pa_{\chi} W$.
These are exactly the flow equations (\ref{flow}) we explained before.
From (\ref{third}), there exists $2 \pa_r |\eta| = -\sqrt{2} g |\eta| W$.
It is straightforward to check that all other supersymmetry parameters 
$\epsilon^i$ where $i=5,6,7,8$ vanish. So we have checked these for
$\varphi =\frac{\pi}{2}$ explicitly.

According to the branching rule of ${\bf 6}$ representation
corresponding to spin $\frac{3}{2}$ field of $SU(4)$
under the $SU(2) \times SU(2)$, ${\bf 6} \rightarrow [({\bf 1}, {\bf
  1}) 
\oplus ({\bf 1}, {\bf 1})]
\oplus ({\bf 2}, {\bf 2})$, the two singlets in square bracket
correspond to the component of massless graviton of the ${\cal N}=4$ theory. 
From the branching rule of ${\bf 15}$ representation
corresponding to spin $1$ field of $SU(4)$
under the $SU(2) \times SU(2)$, ${\bf 15} \rightarrow [({\bf 1}, {\bf
  1})]
\oplus ({\bf 1}, {\bf 3}) \oplus ({\bf 3}, {\bf 1})
\oplus ({\bf 2}, {\bf 2}) \oplus ({\bf 2}, {\bf 2})$, the  singlet 
in square bracket corresponds to the component of the massless graviton 
of the ${\cal N}=4$ theory. 
Finally, spin 2 field with the breaking ${\bf 1} \rightarrow ({\bf 1},
{\bf 1})$
is located at the remaining component of 
${\cal N}=4$ massless graviton multiplet. 

From the decomposition of spin 1 field above, the representation 
$({\bf 1}, {\bf 3}) \oplus ({\bf 3}, {\bf 1})$ corresponds to the
massless vector multiplet of ${\cal N}=4$ theory.
According to the branching rule of ${\bf 10}$ representation
corresponding to spin $\frac{1}{2}$ field of $SU(4)$
under the $SU(2) \times SU(2)$, ${\bf 10}, \overline{{\bf 10}} 
\rightarrow [({\bf 1}, {\bf
  3}) 
\oplus ({\bf 3}, {\bf 1})]
\oplus ({\bf 2}, {\bf 2})$, the representations in square bracket
correspond to the component of massless vector multiplet 
of the ${\cal N}=4$ theory. Moreover, the branching rule for spin zero
field
provides that each representation 
$({\bf 1}, {\bf 3}) \oplus ({\bf 3}, {\bf 1})$
from the two ${\bf 15}$'s of $SU(4)$, as above,
corresponds to the remaining component of massless vector multiplet of
the ${\cal N}=4$ theory. 

\section{The ${\cal N}=4$ supersymmetric 
membrane flows in three dimensions }

Let us contract the $\phi_{IJKL}$ (\ref{phiijkl}) with the gamma matrices $\Gamma^I$.
We use $\Gamma^1={\bf 1}_{8\times 8}$ and $SO(7)$ gamma matrices
$\Gamma^J$
where $J=2, 3, \cdots, 8$ \cite{AW01}. We multiply 
$\left(\begin{array}{cc}
 0 &  {\bf 1}_{4 \times 4} \\
{\bf 1}_{4\times 4}  & 0\\
\end{array}\right)$ with $\Gamma^I$ and  
construct new gamma matrices
$\left(\begin{array}{cc}
 0 &  {\bf 1}_{4 \times 4} \\
{\bf 1}_{4\times 4}  & 0\\
\end{array}\right) \Gamma^I \left(\begin{array}{cc}
 0 &  {\bf 1}_{4 \times 4} \\
{\bf 1}_{4\times 4}  & 0\\
\end{array}\right)$.
Let us define the following quantity which was introduced in \cite{GW}
with new gamma matrices \bea
S_{AB} \equiv \phi_{IJKL} (\Gamma^{IJKL})_{AB},
\nonu
\eea
and one obtains \footnote{For the ${\cal N}=2$ $SU(3) \times U(1)_R$
  invariant supersymmetric flow, one can also obtain 
$S = \mbox{diag} (\la-i \frac{4\la'}{3}, \la + i \frac{4\la'}{3}, \la, \la, \la, \la, 
-3\la, -3\la)$ where $\la$ is a scalar field and $\la'$ is a
pseudo scalar field. 
The supergravity scalar field corresponds to the mass terms of $(77)$
plus $(88)$ components in the boundary theory. 
The supergravity pseudo-scalar field corresponds to the mass terms of $(22)$
minus $(11)$ components in the boundary theory. 
For ${\cal N}=1$ $G_2$ invariant supersymmetric
flow one has $S = \mbox{diag} (\la \cos \alpha-i \frac{7 \la \sin \alpha}{3}, 
\la  \cos \alpha+ i \frac{\la \sin \alpha}{3}, \la \cos \alpha +i \frac{\la \sin
  \alpha}{3}, 
\la \cos \alpha + i \frac{\la \sin \alpha}{3}, \la \cos
\alpha + i \frac{\la \sin \alpha}{3}, \la \cos \alpha + i \frac{\la \sin \alpha}{3}, 
-7 \la \cos \alpha + i \frac{\la \sin \alpha}{3} )$ where $\la$ and
$\al$ are two supergravity fields. In this case, the scalar field 
$\la \cos \alpha$ corresponds to the mass term of $(88)$ component in the boundary 
theory and pseudo-scalar field $\la \sin \alpha$ 
does the mass term of $(11)$ component. }
\bea
S = \mbox{diag} (\al, \al, \al, \al, -\al+i 2\chi, -\al+i 2\chi, 
-\al-i 2\chi, -\al-i 2\chi).
\label{S}
\eea
We want to see how the supergravity scalar fields $(\alpha, \chi)$ map onto
the corresponding boundary field theory object. 

Let us consider the BL theory \cite{BL0711} 
with $SO(4)$ gauge group and matter fields. Although this is not directly
connected to AdS/CFT correspondence in the large $N$ limit due to the
fact that there exist only two M2-branes, 
the approach to BL theory will give us some hints for the gravity duals.
We impose the constraint on the $\epsilon$ parameter that satisfies the 
$\frac{1}{2}$ BPS condition(the number of supersymmetries is eight) \cite{HLL}:
$
\Gamma^{78910}\epsilon = -\epsilon$ with eleven-dimensional gamma matrices. 
The variation for  the bosonic mass term  
\bea
{\cal L}_{b.m.} = -\frac{1}{2} h_{ab} X_I^a  (m^2)_{IJ} X_J^b,
\label{bosonic}
\eea
plus the
fermionic mass term with equal masses 
\bea
{\cal L}_{f.m.}  =  -\frac{i}{2} h_{ab} \bar{\Psi}^a 
\left(     m \Gamma^{4589} - m \Gamma^{45710}  \right) \Psi^b,
\nonu
\eea
in the BL theory leads to \cite{Ahn0812}
\bea
\delta {\cal L}  =   i h_{ab} X_I^a  (m^2)_{IJ} \bar{\Psi}^b \Gamma_J \epsilon  
- i h_{ab} \bar{\Psi}^a  \left(   m \Gamma^{4589} -
  m \Gamma^{45710}  \right)^2
X_I^b \Gamma_I \epsilon. 
\nonu
\eea
Then the bosonic mass term $(m^2)_{IJ} \Gamma_J$
should take the form
$
4m^2 ( \Gamma_7 +\Gamma_8 + \Gamma_9 +\Gamma_{10})$. 
Then the diagonal bosonic mass term has nonzero
component only for ($77,88,99$ and $1010$) and other components($33, 44, 55,
66$) are vanishing. The
degeneracy $4$ is related to the ${\cal N}=4$ supersymmetry.  
Then one obtains the bosonic mass term which 
appears in (\ref{bosonic})
\bea
(m^2)_{IJ} = \diag(0,0,0,0,4m^2,4m^2,4m^2,4m^2).
\nonu
\eea
Of course, there exist the quartic terms for bosonic field $X_I^a$ in
the deformed full Lagrangian. This  can be seen from the component
expansion in ${\cal N}=2$ superspace later.

Then how do we understand this analysis in ${\cal N}=2$ superspace approach?
From the superpotential \cite{BKKS} of the BL theory in ${\cal N}=2$ superspace with 
$SU(2) \times SU(2)=SO(4)$ gauge group, by adding 
the quadratic mass deformation (\ref{bosonic}),
we expect to have  
the full
superpotential \footnote{One cannot conclude that the dimension of
  monomials in the superpotential is the sum of the dimensions of each
component. There is no conformal fixed point in the IR(for the bulk
theory, look at the Figure 1) and the RG flow
does not terminate at a conformal field theory fixed point, contrary
to the case of \cite{Ahn0806n2,BKKS,KKM}.}
\bea 
-\frac{1}{ 8 \cdot 4!} 
 \epsilon_{ABCD} \epsilon^{abcd} {\cal Z}_a^A {\cal Z}_b^B
{\cal Z}_c^C {\cal Z}_d^D -2 m^2  ({\cal Z}_a^3)^2 -2 m^2  ({\cal Z}_b^4)^2,
\label{fullexp}
\eea
where ${\cal Z}_a^A $ \cite{BKKS} is an ${\cal N}=2$ chiral 
superfield with $SU(4)$ index $A=1, 2, \cdots, 4$ and with $SO(4)$ gauge
group index $a=1,2, \cdots, 4$. 
The global symmetry $SO(8)_R$ of the theory
is broken to $[SU(2) \times U(1)]^2$ symmetry.
The mass terms, the last two terms 
in (\ref{fullexp}), break ${\cal N}=8$ down to ${\cal N}=4$.
The mass-deformed theory has matter multiplet in 
two flavors ${\cal Z}^1$ and ${\cal Z}^2$
transforming in the adjoint of the gauge group. 
We'll describe the moduli space characterized by these massless ${\cal
Z}^1$ and ${\cal Z}^2$ in next section.
The $SO(8)_R$ symmetry of the 
${\cal N}=8$ gauge theory is broken to $[SU(2) \times SU(2)]_R \times U(1)^2$
where the $[SU(2) \times SU(2)]_R$ is the R-symmetry of the ${\cal N}=4$ theory
and  one of them $SU(2)$ is a flavor symmetry.
Therefore, we turn on the mass perturbation in the UV and flow to the
IR.
This maps to turning on the scalar fields $\alpha$ and $\chi$ in the $AdS_4$
supergravity where the scalars approach to zero in the UV($r
\rightarrow \infty$) and develop a nontrivial profile as a function of
$r$ (\ref{flow}) and (\ref{equa1})
becoming more significantly different from zero as one goes to the
IR($r \rightarrow -\infty$).
Similar flow but nonsupersymmetric $SU(2) \times SU(2) \times U(1)$ 
critical point was studied in \cite{Ahn0809}.

Motivated by the fact that the two M2-branes theory of BL theory is
equivalent to $U(2) \times U(2)$ Chern-Simons matter theory with level
$k=1$ or $k=2$(there is further enhancement from ${\cal N}=6$ 
to ${\cal N}=8$
supersymmetry and two extra supersymmetries transform as $SO(6)=SU(4)$ 
singlets), 
it is natural to ask what happens for Chern-Simons
matter theory when we turn on mass perturbation in the gauged supergravity?
Let us consider the $U(2) \times U(2)$ Chern-Simons matter theory.
From the superpotential \cite{KKM} of $U(2) \times U(2)$ Chern-Simons
matter theory, the quadratic mass deformations are
added as follows: 
\bea 
\frac{T^{-4}}{4!} 
 \epsilon_{ABCD} \Tr {\cal Z}^A {\cal Z}^{\ddagger B}
{\cal Z}^C {\cal Z}^{\ddagger D} -2 m^2 T^{-2}  \Tr {\cal Z}^3 {\cal
  Z}^{\ddagger 3} -
2 m^2 T^{-2} \Tr {\cal Z}^4 {\cal Z}^{\ddagger 4},
\label{bl}
\eea
where ${\cal Z}^A $ is also an ${\cal N}=2$ chiral 
superfield with $SU(4)$ index $A=1, 2, \cdots, 4$ and an operation
$\ddagger$ is defined by ${\cal Z}^{\ddagger A} \equiv - i \sigma_2
({\cal Z}^A)^T i \sigma_2$. 
The relation of ${\cal Z}^A$ 
to $SO(4)$ notation ${\cal Z}_a^A$ is described in \cite{BKKS}. 
Note that there exist relations 
${\cal Z}^3= {\cal W}_1^{\ddagger}$ and ${\cal Z}^4={\cal
  W}_2^{\ddagger}$.
The $T^2$ in (\ref{bl}) is 
a monopole operator \cite{BKW} which creates two units of magnetic
flux when the level $k=1$ for the $U(1)$ field in the $U(1) \times
U(1)$ Chern-Simons matter theory.  
Note that 
the $SU(4)$ subgroup of $SO(8)$ consists of complex rotation on the
four
component complex $SO(8)$ vector $\vec{v}$. 
Also there exists $U(1)$ subgroup 
which transforms this complex vector as itself 
with an extra overall phase.      
Then the independent two $SU(2)$ transformations, acting on 
the first two components of a complex vector and on 
the last two components of a complex vector respectively
and  the overall phase rotation gives 
a manifest $[SU(2) \times SU(2)]_R \times U(1)$ symmetry. 
When the superpotential $\frac{1}{4!} 
 \epsilon_{ABCD} \Tr {\cal Z}^A {\cal Z}^{\ddagger B}
{\cal Z}^C {\cal Z}^{\ddagger D}$ which is proportional to 
$ \epsilon_{ABCD} \epsilon^{abcd} {\cal Z}_a^A {\cal Z}_b^B
{\cal Z}_c^C {\cal Z}_d^D$ in (\ref{bl}) is written as 
$\frac{1}{4} \epsilon_{AC} \epsilon^{BD} \Tr {\cal Z}^A {\cal W}_B
{\cal Z}^C {\cal W}_D$,  the $[SU(2) \times SU(2)]_R \times U(1)$ 
symmetry is manifest where the baryonic $U(1)$ acts as
${\cal Z}^A \rightarrow e^{\frac{2\pi i}{k}} {\cal Z}^A$ and $
{\cal W}_B \rightarrow e^{-\frac{2\pi i}{k}} {\cal W}_B$ 
with level $k$ \cite{BKKS,KKM}.

In order to see the relation between the deformed Lagrangian from BL
theory
and ${\cal N}=2$ superspace description, we need to integrate the
superpotential over the fermionic coordinates. 
After the integration over the superspace explicitly, 
the quadratic deformation, the last two terms in (\ref{bl}), 
leads to
\bea
&& m^2 T^{-2} \int d^2 \theta   \Tr  (
{\cal Z}^3 {\cal Z}^{\ddagger 3} +{\cal Z}^4 {\cal Z}^{\ddagger 4} )
\nonu \\
&&  =  - m^2 T^{-2} \Tr (\zeta^3 \zeta^{\ddagger 3} + 
\zeta^4 \zeta^{\ddagger 4}) + 2m^2 T^{-2} \Tr ( F^3
 Z^{\ddagger 3} +  F^4
 Z^{\ddagger 4}) \nonu \\ 
&& = - m^2 T^{-2} \Tr  ( \zeta^3 \zeta^{\ddagger 3} +
\zeta^4 \zeta^{\ddagger 4} ) + T^{-4} \frac{m^2 L}{3} ( 
\epsilon^{ABC} \Tr \bar{Z}_A \bar{Z}_B^{\ddagger} \bar{Z}_C
Z^{\ddagger 3} +
\epsilon^{\hat{A}\hat{B}\hat{C}} \Tr \bar{Z}_{\hat{A}} 
\bar{Z}_{\hat{B}}^{\ddagger} \bar{Z}_{\hat{C}}
Z^{\ddagger 4}),
\nonu
\eea
where the auxiliary field $F$ is replaced using the equation of motion
in the last line:$F^A = -\frac{L}{6} \epsilon^{ABCD} \bar{Z}_B
\bar{Z}_C^{\ddagger} 
\bar{Z}_D$ \cite{BKKS} and the component expansions for ${\cal N}=2$
superfields are used. 
Here the $L$ is a coefficient of the superpotential term
in ${\cal N}=2$ superspace. 
The mass terms for the fermions correspond to the action of $S$
introduced in (\ref{S}) where $\zeta^3$ is related to a complex
combination of 5-th component $2\chi$ and 7-th component $-2\chi$ 
of (\ref{S})
and $\zeta^4$ is related to a complex combination of 6-th component $2\chi$ 
and 8-th component $-2\chi$.  
Note that in component expansion, there exist  quartic terms 
as well as the mass terms for the fermions.

There are also similar constructions \cite{N4,IK,HLLLP} which have ${\cal N}=4$
supersymmetry. It would be interesting to see how they are related to 
the present mass-deformed BL or Chern-Simons matter theories.  
 
\section{The M2-brane probe }

Now we want to understand the eleven-dimensional solution from the
scalar field configuration.
By applying the formula of \cite{dWNW} to the ${\cal N}=4$ gauged supergravity,
the 11-dimensional metric is given by \cite{GW} 
\bea
ds_{11}^2 & = &
\Omega^2
\left( e^{2A(r)} dx_{\mu}^2+dr^2 \right)+   \frac{4L^2 
\Omega^2}{c} d \theta^2
+  L^2 \Omega^2 \rho^2 \cos^2 \theta
\left( \frac{1}{X_1}  \si_1^2 + \frac{1}{X_1}  
\si_2^2 + \frac{1}{c X_2} \si_3^2 \right) \nonu \\
&+&
 L^2 \Omega^2  \sin^2 \theta 
\left( \frac{1}{X_2} \tau_1^2 + \frac{1}{X_2} \tau_2^2 +
\frac{1}{c X_1} \tau_3^2 \right),
\label{11dmetricn4}
\eea
where we introduce the following quantities
\bea
c & \equiv & \cosh (2 \chi), \qquad 
\rho \equiv e^{\alpha}, \qquad
 X_1 
\equiv \cos^2 \theta + \rho^2  \sin^2 \theta \cosh (2\chi),
\nonu \\ 
X_2 & \equiv & 
\cos^2 \theta \cosh (2\chi) + \rho^2 \sin^2 \theta,
\qquad
\Omega 
\equiv \left[ \frac{X_1 X_2 \cosh (2\chi)}{\rho^2} \right]^{\frac{1}{6}}.
\label{relations}
\eea
The $\si_i$ and $\tau_i$ are independent sets of left-invariant 
one-forms on $SU(2)$. This metric (\ref{11dmetricn4}) 
has a manifest symmetry of
$SU(2)_{\si} \times U(1)_{\si}$ and $SU(2)_{\tau} \times 
U(1)_{\tau}$.
The $U(1)_{\si}$ rotates $\si_1$ into $\si_2$ while 
the $U(1)_{\tau}$ rotates $\tau_1$ into $\tau_2$.
The Killing spinors are singlets under $SU(2)_{\tau} \times
U(1)_{\sigma}$
and transform as ${\bf 2}_{\pm 1}$ under $SU(2)_{\sigma} \times
U(1)_{\tau}$.
We'll see that $SU(2)_{\tau}$ acts on the complex structure in the
hyper-Kahler moduli space. 
Note that the additional isometries of the metric transverse to the
M2-brane amount to additional global symmetries of boundary field
theory on the M2-brane.

The 3-form potential \cite{GW} takes the form of 
\bea
A^{(3)} &=& \widetilde{W} e^{3A} \; dt \wedge dx_1 \wedge dx_2 +
L^3 \tanh (2\chi)
 \sin \theta \; \cos \theta \; d \theta \wedge 
\si_3 \wedge \tau_3 \nonu \\
&+&  \frac{L^3 \rho^2} {2X_1} \sinh (2\chi) \; \sin^2 \theta 
\; \cos^2 \theta \; \si_1 \wedge
\si_2 \wedge \tau_3  \nonu \\
&+& \frac{ L^3}{2 X_2} \sinh (2\chi) 
\; \sin^2 \theta \; \cos^2 \theta
\; \si_3 \wedge \tau_1 \wedge \tau_2.
\label{3formpot}
\eea
Here $\widetilde{W}$ 
is called ``geometric'' superpotential \cite{KW}
and it is given by 
\bea
\widetilde{W} & = & \frac{ X_1}{2\rho} =
\frac{1}{2 \rho} \left[
\cos^2 \theta + \rho^2  \sin^2 \theta \cosh (2\chi) \right],
\label{tildeW}
\eea
which is exactly the same as a half of the superpotential (\ref{superpot})
\bea
\widetilde{W}=\frac{W}{2},
\nonu
\eea
at $\theta=\frac{\pi}{4}$. 

Let us describe the ${\cal N}=8$ four-dimensional gauged supergravity.
Now we go to the $SL(8,{\bf R})$ basis \cite{KW} and introduce the rotated vielbeins 
\bea
U^{ij}_{\,\,\,\,IJ} & = & u^{ij}_{\,\,\, ab} (\Gamma_{IJ})^{ab}, 
\qquad
V^{ijIJ} = v^{ijab} (\Gamma_{IJ})^{ab},
\nonu \\
U_{ij}^{\,\,\,\,IJ} & = & u_{ij}^{\,\,\, ab} (\Gamma_{IJ})^{ab}, 
\qquad
V_{ijIJ} = v_{ijab} (\Gamma_{IJ})^{ab},
\nonu
\eea
where all indices $i, j$ and $a, b$ run from 1 to 8 and 
correspond to the realization of $E_{7(7)}$ in the $SU(8)$ basis
and $\Gamma_{IJ}$ are the $SO(8)$ generators in \cite{AI}.
We also define the following quantities
\bea
A_{ijIJ} & = & 
\frac{1}{\sqrt{2}} \left( U_{ij}^{\,\,\,\,IJ} + V_{ijIJ}
\right), 
\qquad
B_{ij}^{\,\,\,IJ} =
\frac{1}{\sqrt{2}} \left( U_{ij}^{\,\,\,\,IJ} - V_{ijIJ} \right),
\nonu \\
C^{ij}_{\,\,\,IJ} & = & \frac{1}{\sqrt{2}} \left(U^{ij}_{\,\,\,\,IJ} +
V^{ijIJ}  \right), \qquad 
D^{ijIJ} =\frac{1}{\sqrt{2}} \left(-U^{ij}_{\,\,\,\,IJ} +
V^{ijIJ}  \right).
\label{ABCD}
\eea
The full expressions for these are in (\ref{A}), (\ref{redef}), (\ref{B}), 
(\ref{C}) and (\ref{D}) from 
the appendix B.

Then the ``geometric'' T tensor \cite{KW} can be written as
\bea
\widetilde{T}_l^{\,kij} = \frac{1}{21\sqrt{2}} C^{ij}_{\,\,LM} 
\left( A_{lmJK} D^{kmKI} \, \delta^L_{\,I} \, x_M x_J -B_{lm}^{\,\,\,JK} 
C^{km}_{\,\,\,KI} \, \delta^M_{\,J} \, x_L x_I \right),
\label{ttilde}
\eea
where we have a relation between $x_I$ and 
$Y_I$ that is a coordinate for ${\bf R}^8$ where $\sum_{I=1}^{8} (Y_I)^2
=1$ and that 
is introduced in order to absorb the cross  terms between $x_I$'s:
\bea
Y_1 & \equiv & \frac{1}{\sqrt{2}} (x_2-x_6), \qquad Y_2
\equiv -\frac{1}{\sqrt{2}} (x_3-x_7),
\nonu \\
Y_3 & \equiv & \frac{1}{\sqrt{2}} (x_4-x_8), \qquad Y_4
\equiv -\frac{1}{\sqrt{2}} (x_1-x_5),
\nonu \\
Y_5 & \equiv & \frac{1}{\sqrt{2}} (x_2+x_6), \qquad Y_6
\equiv \frac{1}{\sqrt{2}} (x_3+x_7),
\nonu \\
Y_7 & \equiv & \frac{1}{\sqrt{2}} (x_4+x_8), \qquad Y_8
\equiv \frac{1}{\sqrt{2}} (x_1+x_5).
\nonu 
\eea
From this, the corresponding ``geometric'' $\widetilde{A}_1$ tensor is given by
$
\widetilde{A}_1^{\,ij} = \widetilde{T}_m^{\,\,imj}$ and we present
them in (\ref{tilone}), (\ref{tiltwo}), (\ref{tilthree}),
(\ref{tilfour}), 
(\ref{tilfive}) from appendix B.
The idea of \cite{KW} is to introduce geometric analogue of the
T-tensor by replacing $\delta^{IJ}$ by $x^I x^J$. 

By computing the real part of $11$-component of the $\widetilde{A}_1$ 
tensor (\ref{tilone}) with $\phi=0$ and $\varphi=\frac{\pi}{2}$, one obtains 
the geometrical superpotential $W_{gs}$ as follows:
\bea
W_{gs}=e^{-\alpha} \left( Y_1^2 +Y_2^2+Y_3^2+Y_4^2\right) 
+ e^{\alpha} \cosh (2\chi) \left( Y_5^2+Y_6^2+Y_7^2+Y_8^2 \right).
\label{geosuperpot}
\eea
Note that half of the transverse coordinates has common factor and 
the remaining half of them has other common factor. We want to see how
this geometric superpotential is related to the superpotential 
(\ref{superpot}) or (\ref{tildeW}). There is a chance to compare this 
(\ref{geosuperpot}) with (\ref{tildeW}) by looking at the $\alpha$ and
$\chi$ dependence.
Through the relations
\bea
Y_1 & = &  \cos \theta \cos (\frac{\alpha_1}{2})
\cos (\frac{\alpha_2 +\alpha_3}{2}), \qquad 
Y_2 =   \cos \theta \cos (\frac{\alpha_1}{2})
\sin (\frac{\alpha_2 +\alpha_3}{2}),
\nonu \\
Y_3 & = &  \cos \theta \sin (\frac{\alpha_1}{2})
\cos (\frac{\alpha_2 -\alpha_3}{2}), \qquad 
Y_4 =   \cos \theta \sin (\frac{\alpha_1}{2})
\sin (\frac{\alpha_2 -\alpha_3}{2}),
\nonu \\
Y_5 & = &  \sin \theta \cos (\frac{\beta_1}{2})
\cos (\frac{\beta_2 +\beta_3}{2}), \qquad 
Y_6 =   \sin \theta \cos (\frac{\beta_1}{2})
\sin (\frac{\beta_2 +\beta_3}{2}),
\nonu \\
Y_7 & = &  \sin \theta \sin (\frac{\beta_1}{2})
\cos (\frac{\beta_2 -\beta_3}{2}), \qquad 
Y_8 =   \sin \theta \sin (\frac{\beta_1}{2})
\sin (\frac{\beta_2 -\beta_3}{2}),
\nonu 
\eea
this geometric superpotential (\ref{geosuperpot}) 
reduces to 
\bea
W_{gs} =2\widetilde{W},
\nonu
\eea 
with (\ref{tildeW}). Here $\alpha_i$ and $\beta_i$ are Euler angles
that parametrize two independent sets of $SU(2)$ 
left-invariant one-forms 
respectively. Also the usual spherical coordinates for three-sphere 
can be used to write any element of $SU(2)$.
This is consistent with the definition of geometric superpotential 
through the superpotential 
$8 W =  \frac{1}{2} \sum_{I=1}^8 
\frac{\pa^2}{\pa Y^I \pa Y^I} W_{gs}$ \cite{KW}.

Furthermore, by calculating the $\widetilde{A}_2$ tensor(we do not
present in this paper) obtained from the
geometric T tensor, one arrives at the geometric scalar potential eventually
\bea
V_{gp}(\al,\chi,\theta) & = & 
-g^2 \left( \frac{3}{4} \left| \widetilde{A}_1^{\;ij} \right|^2-\frac{1}{24} \left|
\widetilde{A}^{\;\;i}_{2\;\;jkl}\right|^2 \right) \nonu \\
&=&
\frac{g^2}{4} \cosh (2\chi) \left[ (3+ \cos [4\theta]) \cosh
  [2(\al-\chi)]+ (3+ \cos [4\theta]) \cosh
  [2(\al+\chi)] \right. \nonu \\
& + & \left.  (3+ \cos [4\chi]) \sin^2 (2\theta) - 8 \cos (2\theta) \cosh
  (2\chi) \sinh (2\alpha)
\right].  
\nonu
\eea
This has a critical point at $\alpha=0=\chi$ and $\theta =\frac{\pi}{4}$.

The potential seen by the M2-brane probe \cite{JLP00,KW,JLP01} 
has a factor
\bea
e^{3A} (\Omega^3 - 2\widetilde{W}).
\label{potential}
\eea
The moduli spaces of the brane probe are given by the loci
where the potential vanishes. One sees that 
this potential (\ref{potential}) vanishes at 
\bea
\cos \theta =0 \rightarrow \theta =\frac{\pi}{2},
\nonu
\eea 
where we use (\ref{relations}) and (\ref{tildeW}) or (\ref{geosuperpot}).
On this subspace,  the four-dimensional moduli space
from (\ref{11dmetricn4}), by multiplying the factor $e^A \Omega^{-\frac{1}{2}}$ into
the seven-dimensional internal metric, 
is given by
\bea
ds^2|_{\rm{moduli}}=
 \rho\cosh\left(2\chi\right)e^{A}L^{2}
\left[\frac{1}{L^{2}}dr^{2}+\frac{1}{\rho^{2}}
\left(\tau_{1}^{2}+\tau_{2}^{2}\right)+
\frac{1}{\rho^{2}\cosh^{2}\left(2\chi\right)}\tau_{3}^{2}\right].
\label{mod}
\eea
It is evident that the explicit one $SU(2)$ 
from $[SU(2) \times SU(2)]_R$ invariance is that of the
flavor symmetry. 
Note that the other $SU(2)$ acts on the geometry transverse to both 
the M2-branes and moduli space.
The vacuum expectation values of the massless scalars
are denoted by 
$Z^1$ and $Z^2$(that are $\theta$-independent components of
${\cal Z}^1$ and ${\cal Z}^2$ respectively) 
and $Z^i$'s transform in the fundamental
representation of $SU(2)$ and their complex conjugates transform 
in the anti-fundamental representation of $SU(2)$.
That is, ${\cal Z}^1$ and ${\cal Z}^2$ parametrize the Cartesian coordinates 
$x^3 \equiv r, x^8, x^9, x^{10}$ while  
${\cal Z}^3$ and ${\cal Z}^4$ parametrize the remaining Cartesian coordinates 
$x^4, x^5, x^6, x^7$. 
The $SU(2)$ flavor symmetry implies that 
the Kahler potential is a function of $u^2$ where
let us define 
$
u^2 \equiv Z^1 \bar{Z}_1 + Z^2 \bar{Z}_2$.
Specifying the coordinates and indices into holomorphic and
anti-holomorphic
and if the Kahler structure exists, then the metric is given by 
$ds^2 =\pa_{\mu} \pa_{\bar{\nu}} K(u^2) d Z^{\mu} d Z^{\bar{\nu}}$.
Then the metric turns out to be \cite{JLP01}
\bea
ds^{2}=(K^{'}+u^{2}K^{''})du^{2}+
u^{2}\left[ K^{'}(\tau_{1}^{2}+
\tau_{2}^{2})+(K^{'}+u^{2}K^{''})\tau_{3}^{2}\right].
\label{kahler}
\eea

Is there any connection between (\ref{mod}) and (\ref{kahler})?
Let us compare (\ref{mod}) with (\ref{kahler}). Then we get
\bea
(K^{'}+u^{2}K^{''})du^{2} & = &
\rho\cosh (2\chi)e^{A}dr^{2}, \nonu \\
u^{2}(K^{'}+u^{2}K^{''}) & = & 
\frac{L^{2}e^{A}}{\rho\cosh (2\chi)}, \nonu \\
u^{2}K^{'} & = & 
\frac{L^{2}\cosh (2\chi)e^{A}}{\rho}.
\label{rel}
\eea
From the last equation of (\ref{rel}), 
\bea
u^{2}K^{'}  =  
u^{2}\frac{dK}{d (u^{2})} 
=
\frac{L^{2}\cosh (2\chi) e^{A}}{\rho}.
\label{rell}
\eea
We need to check the second equation of (\ref{rel}).
Since
\bea
u^{2}(K^{'}+u^{2}K^{''}) & = & 
u^{2}\frac{d}{d (u^{2})} (u^{2}K^{'}),
\label{relll}
\eea
by substituting (\ref{rell}) into (\ref{relll}), one has
\bea
u^{2}\frac{d}{d (u^{2})} (u^{2}K^{'}) & = & 
\frac{L}{2\rho\cosh (2\chi)}\frac{d}{dr}
\left(\frac{L^{2}\cosh\left(2\chi\right)e^{A}}{\rho}\right).
\nonu
\eea
Here we use the following change of variable which can be obtained by
using the first and second equations of (\ref{rel})
$
u^{2}\frac{d}{d\left(u^{2}\right)}  =  
\frac{L}{2\rho\cosh\left(2\chi\right)}\frac{d}{dr}$.
Now after using the flow equations (\ref{flow}) explicitly
we obtain
\bea
\frac{L}{2\rho\cosh (2\chi)}\frac{d}{dr}\left(\frac{L^{2}\cosh
(2\chi) e^{A}}{\rho}\right)  =  
\frac{2L^{2}e^{A}}{\rho\cosh (2\chi)},
\nonu
\eea
which is {\it not} the right hand side of second equation (\ref{rel}).
Therefore, 
the metric (\ref{mod}) cannot be written as 
(\ref{kahler}) due to the coefficient one of $\tau_3^2$ in
(\ref{mod}): shortened by a factor $\frac{1}{2}$.
If the coefficient of $\tau_3^2$ in (\ref{mod}) is two, then 
one can express the metric in terms of Kahler potential through 
(\ref{kahler}).

On the other hand, one can write the metric (\ref{mod}) in terms of 
$\chi$ variable instead of $r$ using the flow equation (\ref{flow}) 
as follows \cite{GW}:
\bea
ds^2|_{\rm{moduli}} = \frac{\cosh(2\chi)}{\sinh^3 (2\chi)} d\chi^2 +
\frac{\cosh(2\chi)}{4\sinh(2\chi)}(\tau_1^2 +\tau_2^2) +
\frac{1}{4 \cosh(2\chi) \sinh(2\chi)} \tau_3^2.
\label{modd}
\eea
There is no $\rho$ or $\alpha$ dependence on this moduli space metric.
As $\chi \rightarrow \infty$, the 2-sphere by $\tau_1$ and $\tau_2$
limits a sphere of radius $\frac{1}{2}$.
Moreover, by introducing a new radial variable $\mu$,
the metric (\ref{modd}) can be written as \cite{GW}
\bea
ds^2|_{\rm{moduli}} = \frac{d\mu^2}{ 
\left( 1- \frac{1}{\mu^4} \right)} +
\frac{\mu^2}{4}(\tau_1^2 +\tau_2^2) +
\frac{\mu^2}{4} \left( 1- \frac{1}{\mu^4} \right) \tau_3^2, 
\qquad \mu \equiv \sqrt{\frac{\cosh(2\chi)}{\sinh (2\chi)}}.
\nonu
\eea
As $\chi \rightarrow \infty$ or $\mu \rightarrow 1$, the branes 
are spread out over finite two sphere. 
In this parametrization one can do similar analysis of 
(\ref{kahler})-(\ref{relll}) and obtains the same result: the
difference in the coefficient of the metric. 
For $\theta =\frac{\pi}{2}$, 
the three-form potential (\ref{3formpot}) becomes \cite{GW}
\bea
A^{(3)} \sim H^{-1} dt \wedge d x_1 \wedge d x_2, \qquad H \equiv 
e^{-3A} \Omega^{-3}|_{\theta=\frac{\pi}{2}} = \frac{\sinh^2 
(2\chi)}{\rho^4 \cosh(2\chi)}.
\nonu
\eea
One can easily check that the behavior of the function $H$ gives the
right asymptotics for the uniform distribution of branes spread over
the two sphere at $e^{-2\chi} =0$ which is another good radial coordinate. 

\section{The holographic ${\cal N}=8$ flow 
in four dimensions}

The invariant scalar manifold by Pope-Warner \cite{PW} 
consisting of two scalars 
has a factor $SL(2,{\bf R})$ in $E_{7(7)}$ which has a
maximal non-compact subgroup $SL(8,{\bf R})$. 
Then the noncompact generators of $SL(2, {\bf R})$ 
can be written as
\cite{PW}
\bea
\phi_{ijkl} = (\alpha \cos \zeta + i \alpha \sin \zeta)
\delta_{ijkl}^{1234} +(\alpha \cos \zeta - i \alpha \sin \zeta)
\delta_{ijkl}^{5678}. 
\nonu
\eea
This can be obtained from (\ref{phiijkl}) by taking 
$\phi \rightarrow \zeta$ and $\chi \rightarrow 0$.
See also \cite{AW01,AW02} for the $SO(p) \times SO(8-p)$ invariant
generator of $SL(8, {\bf R})$.
As we did before, the 56-beins ${\cal V}(x)$
can be written as (\ref{calV}) or (\ref{56bein})
and 28-beins $u$ and $v$ 
are summarized in (\ref{res1}) in the appendix C.
It turns out that 
$A_1$ tensor has one real
eigenvalue, $z_1$ 
with degeneracies 8 and has the following form
\bea
A_1 =\mbox{diag} \left(z_1, z_1, z_1, z_1, z_1,
z_1, z_1, z_1 \right), 
\nonu
\eea
with $
z_1  =  \cosh \al$.

The scalar potential
can be written, by combining all the components of
$A_1, A_2$ tensors where (\ref{res2}) gives the data of $A_2$ tensor,  
as
\bea
V(\al)= 
- 2g^2 \left[ 2+ \cosh (2\alpha) \right] =
2g^2 \left[  \left(\frac{\partial
W}{\partial \al}\right)^2  - 3  W^2 \right].
\label{pot1}
\eea
The superpotential is given by
\bea
W(\al) = \cosh \al.
\label{superpot1}
\eea
It turns out that the $A_1$ tensor that appears in the
gravitino transformation rule has $\cosh \alpha$
with degeneracy 8. This implies maximal supersymmetry(${\cal N}=8$).
We'll check this explicitly later.
The scalar potential at the $SO(8)$ UV critical point where $\alpha=0$ 
(or $r \rightarrow \infty$) becomes 
$V=-6g^2$ where $g=\frac{1}{\sqrt{2} L}$ and the superpotential 
$W$ becomes 1.
Note that the supergravity scalar potential (\ref{pot1}) is independent of the
angle $\zeta$.
This critical point is common to both a scalar 
potential (\ref{pot1})
and a superpotential (\ref{superpot1}) and is depicted in Figure 2.

\begin{figure}
\begin{center}
\begin{minipage}[t]{6.8cm}
\centerline{\hbox{\psfig{file=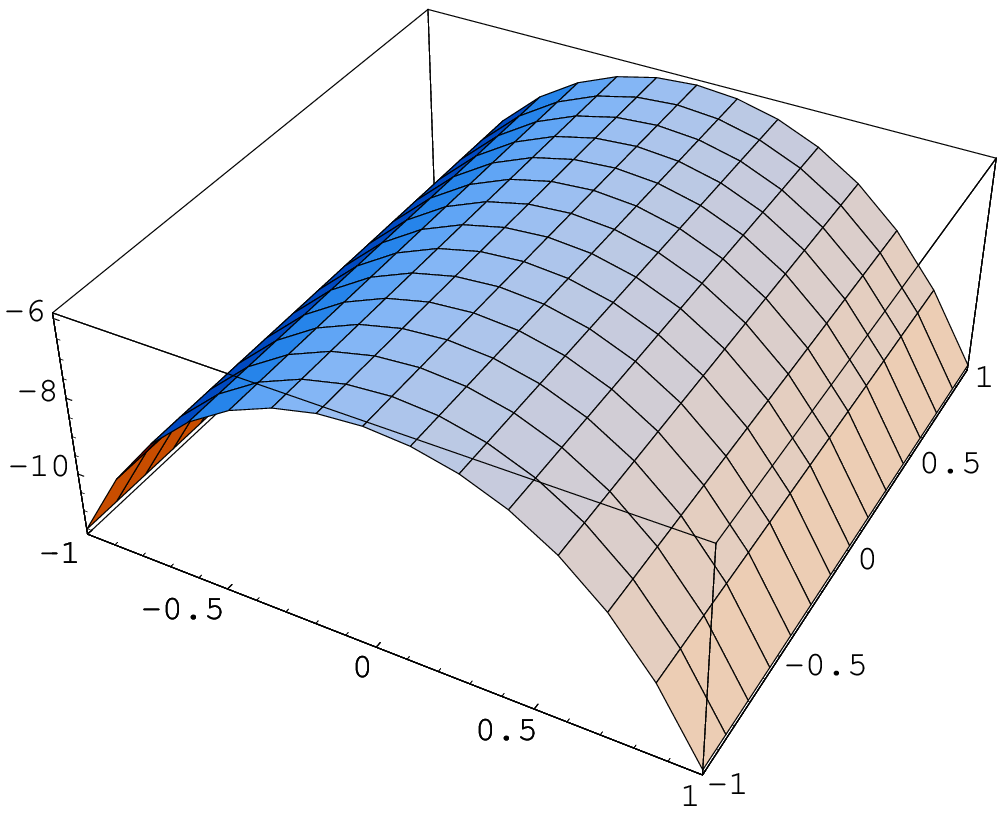,width=6.8cm}}}
\end{minipage}\hspace{1cm}
\begin{minipage}[t]{6.8cm}
\centerline{\hbox{\psfig{file=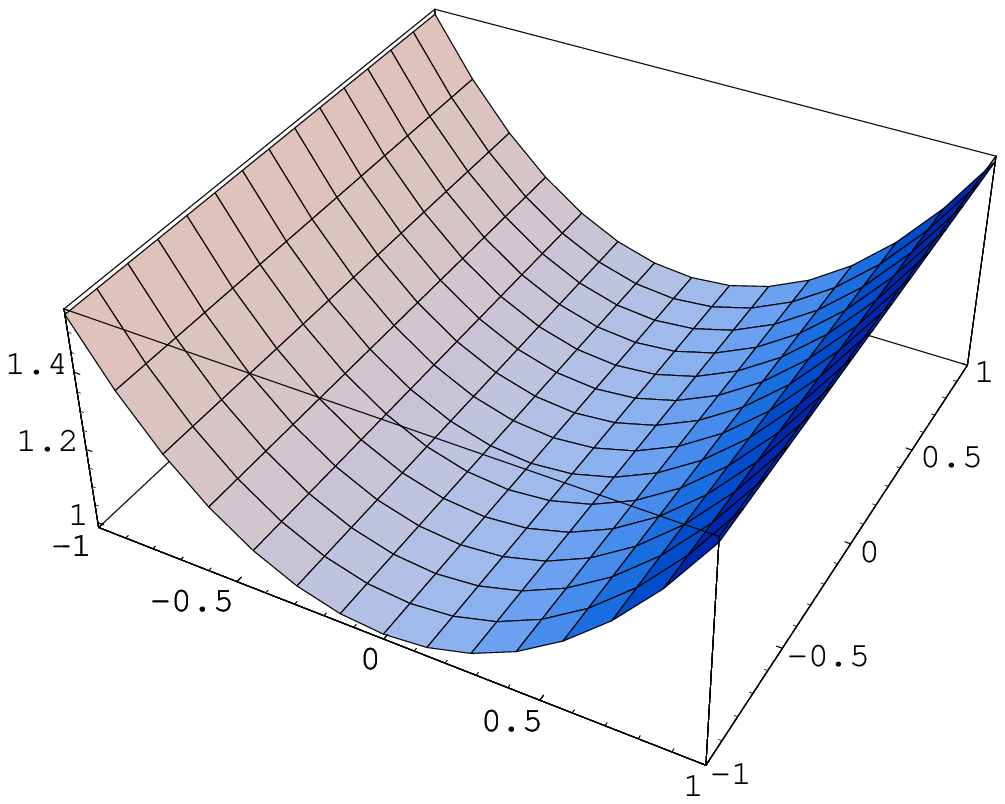,width=6.8cm}}}
\end{minipage}
\caption{\sl Three dimensional plots of 
the scalar potential $V=-2g^2 \left( \cosh 2\alpha +2
 \right) $(left)  in (\ref{pot1}) and 
the superpotential $W=\cosh \alpha$(right) given in (\ref{superpot1}) 
in the 4-dimensional gauged supergravity. 
The axes $(\alpha, \zeta)$ are two vevs that 
parametrize the $SO(4) \times SO(4)$ 
invariant manifold in the 28-beins of the theory. 
The $SO(8)$ UV critical point is located at $\alpha=\zeta=0$.
We have set the $SO(8)$ gauge coupling $g$ in the scalar potential as 
$g=1$.   }
\end{center}
\end{figure}

The resulting Lagrangian of the 
scalar-gravity sector can be obtained 
by finding out the scalar kinetic term.
We arrive at the following scalar kinetic term 
$
(\pa_\mu \alpha)^2$ from (\ref{res3}).
By substituting the domain-wall ansatz 
into the Lagrangian, the 
energy-density per unit area transverse to $r$-direction is obtained.
Then $E[\al, \chi]$ is extremized by the 
BPS domain-wall solutions. 
The maximally supersymmetric flow satisfying the 
first order differential equations for given superpotential 
leads to the solutions for $\alpha(r), A(r)$ and $\zeta(r)={\rm const}$ 
explicitly. 
The dilaton and the axion are function of $\alpha(r)$
and $\zeta$ which remains fixed along the flow: 
\bea  
\frac{d \al}{d r}  = -
\sqrt{2} \, g \, \partial_{\al} W ,\qquad
\frac{d \zeta}{d r}   =    0, \qquad
\frac{d A}{d r}  =   \sqrt{2} \, g \, W.
\label{flow1}
\eea
The solutions for flow equations are given by 
\bea
e^{\alpha}=\coth \left(\frac{r}{2L}\right), \qquad
e^{A}=\sinh \left(\frac{r}{L} \right)=
\frac{1}{\sinh \alpha}, \qquad 
\zeta={\rm const}.
\nonu
\eea 
Note that the flow is maximally supersymmetric for all choices of
$\zeta$
and the potential (\ref{pot1}), superpotential (\ref{superpot1}) and 
the flow (\ref{flow1}) do not depend on $\zeta$.

The vanishing of variation of  $\chi^{ijk}$ for supersymmetry
parameter $\epsilon^i$ where $i=1, \cdots, 8$ 
leads to 
\bea
\sqrt{2}  e^{i\zeta} \left[ \partial_{\mu}\alpha +\frac{i}{2}
\sinh\left(2\alpha\right) \partial_{\mu}\zeta \right] 
\left(
\begin{array}{c}
\eta_2^{\ast}  \\
\eta_1^{\ast}  \\
\end{array}
\right)
& = &
-2 g  e^{i\zeta} \frac{\partial W}{\partial\alpha}
\left( \begin{array}{c}
\eta_1  \\
\eta_2  \\
\end{array}
\right),  \nonu \\
\sqrt{2}  e^{-i\zeta} \left[ \partial_{\mu}\alpha - \frac{i}{2}
\sinh\left(2\alpha\right) \partial_{\mu}\zeta \right] 
\left(
\begin{array}{c}
\eta_2^{\ast} \\
\eta_1^{\ast}  \\
\end{array}
\right) 
& = &
-2g e^{-i\zeta} \frac{\partial W}{\partial\alpha}
\left(\begin{array}{c}
\eta_1  \\
\eta_2  \\
\end{array}
\right).
\label{equa}
\eea
By multiplying the $\eta_2$ into the
first row and $\eta_1$ into the second row in the first equation 
(\ref{equa}) and
combining these two relations, one obtains
$|\eta_1(r)|=|\eta_2(r)|\equiv |\eta(r)|$.
Let us consider the equation 
\bea
2\partial_{r}  \left(\begin{array}{c}
\eta_1 \\
\eta_2 \\
\end{array}
\right) = \pm i \partial_{\mu}\zeta
\sinh^{2} \alpha  \left(\begin{array}{c}
\eta_1 \\
\eta_2 \\
\end{array}
\right)
-\sqrt{2}g W  \left(
\begin{array}{c}
\eta_2^{\ast}  \\
\eta_1^{\ast} \\
\end{array}
\right),
\nonu
\eea
where 
the upper sign is on the supersymmetry parameter 
$\epsilon^{i=1,2,3,4}$ and the lower sign is on 
 the supersymmetry parameter $\epsilon^{i=5,6,7,8}$
and substitute the $\eta_1$
and $\eta_2$. 
We used the result (\ref{calB1}) in the appendix C.
By multiplying $e^{-i\beta}$ into the first row and 
$e^{-i\delta}$ into the second row and subtracting these two, then 
one obtains $(\pa_r \beta -\pa_r \delta)\eta$ which implies that 
$\beta = \delta + \mbox{const}$.   
Now let us analyze the equation 
\bea
\partial_{r}A \left(\begin{array}{c}
\eta_1  \\
\eta_2 \\
\end{array}
\right)  =  \sqrt{2}g W \left(
\begin{array}{c}
\eta_2^{\ast}  \\
\eta_1^{\ast} \\
\end{array}
\right).
\nonu
\eea
From the first row, 
one gets $\pa_r A = \sqrt{2} g W e^{-2i\beta}$. This implies that 
$W e^{-2i\beta}$ should be real because $\pa_r A$ is real.

Let us compare the first and second equations of (\ref{equa}).
Then we obtain $e^{2i\beta} \frac{\partial 
W}{\partial\alpha} = e^{-2i\beta} 
\frac{\partial W}{\partial\alpha}$ because the left hand side 
of second equation of (\ref{equa}) is a complex conjugation of the
left hand side of first equation. This leads to the fact that 
$\beta=0$. Therefore the expression of the
bracket in (\ref{equa}) should be real. So one has $\pa_r \zeta =0$.
By collecting all the informations so far, one obtains 
the flow equations (\ref{flow1}) we explained before.
We have checked that all the supersymmetry parameters 
$\epsilon^i$ where $i=1, \cdots, 8$ do not vanish explicitly.

According to the branching rule \cite{ps,PS1,PS2} of ${\bf 8}_s$ representation
corresponding to spin $\frac{3}{2}$ field of $SO(8)$
under the $[SO(4)]^2 =[SU(2)]^4$, ${\bf 8}_s \rightarrow ({\bf 1},
{\bf 2}, {\bf
  1}, {\bf 2}) 
\oplus ({\bf 2}, {\bf 1}, {\bf 2}, {\bf 1})$, the two singlets 
correspond to the component of massless graviton of the ${\cal N}=8$ theory. 
From the branching rule of ${\bf 28}$ representation
corresponding to spin $1$ field of $SO(8)$
under the $[SU(2)]^4$, ${\bf 28} \rightarrow ({\bf 2}, {\bf 2}, {\bf 2}, {\bf
  2})
\oplus ({\bf 1}, {\bf 3}, {\bf 1}, {\bf 1}) \oplus [({\bf 1}, {\bf 1},
{\bf 1}, {\bf 3})]
\oplus ({\bf 1}, {\bf 1}, {\bf 3}, {\bf 1}) \oplus ({\bf 3}, {\bf 1},
{\bf 1}, {\bf 1})$, the  singlet in square bracket corresponds 
to the component of the massless graviton 
of the ${\cal N}=8$ theory. 
Finally, spin 2 field with the breaking ${\bf 1} \rightarrow ({\bf 1},
{\bf 1}, {\bf 1}, {\bf 1})$
is located at the remaining component of 
${\cal N}=8$ massless graviton multiplet. 

From the decomposition of spin 1 field above, the representation 
$({\bf 1}, {\bf 1}, {\bf 3}, {\bf 1}) \oplus ({\bf 1}, {\bf 3}, {\bf
  1}, 
{\bf 1}) \oplus ({\bf 3}, {\bf 1}, {\bf 1}, {\bf 1})$ correspond to the
massless vector multiplet of ${\cal N}=8$ theory.
According to the branching rule of ${\bf 56}_s$ representation
corresponding to spin $\frac{1}{2}$ field of $SO(8)$
under the $[SU(2)]^4$, ${\bf 56}_s 
\rightarrow [({\bf 2}, {\bf
  3}, {\bf 2}, {\bf 1}) 
\oplus ({\bf 1}, {\bf 2}, {\bf 3}, {\bf 2}) \oplus ({\bf 3},{\bf 2},{\bf 1},{\bf 2})]
\oplus ({\bf 2}, {\bf 1}, {\bf 2}, {\bf 3}) \oplus ({\bf 1}, {\bf 2},
{\bf 1}, {\bf 2}) \oplus ({\bf 2}, {\bf 1}, {\bf 2}, {\bf 1})$, 
the subset of the representations in square bracket
corresponds to the component of massless vector multiplet 
of the ${\cal N}=8$ theory. Moreover, the branching rules for spin zero
field
${\bf 35}_v \rightarrow ({\bf 3},{\bf 3},{\bf 1},{\bf 1} ) \oplus
({\bf 2},{\bf 2},{\bf 2},{\bf 2}) \oplus
({\bf 1},{\bf 1},{\bf 3},{\bf 3}) 
\oplus ({\bf 1},{\bf 1},{\bf 1},{\bf 1})$ and 
${\bf 35}_c \rightarrow ({\bf 1},{\bf 3},{\bf 3},{\bf 1} ) \oplus
({\bf 2},{\bf 2},{\bf 2},{\bf 2}) \oplus
({\bf 3},{\bf 1},{\bf 1},{\bf 3}) 
\oplus ({\bf 1},{\bf 1},{\bf 1},{\bf 1})$ 
provide the remaining component of massless vector multiplet of
the ${\cal N}=8$ theory.

\section{The ${\cal N}=8$ supersymmetric 
membrane flows in three dimensions }

Let us consider the BL theory with $SO(4)$ gauge group and matter fields.
The variation for  the bosonic mass term  
(\ref{bosonic})
plus the
fermionic mass term  
\bea
{\cal L}_{f.m.}  =  -\frac{i}{2} h_{ab} \bar{\Psi}^a 
     m \Gamma^{3456} \Psi^b,
\nonu
\eea
leads to the following variation
\bea
\delta {\cal L}  =   i h_{ab} X_I^a  (m^2)_{IJ} \bar{\Psi}^b \Gamma_J \epsilon  
- i h_{ab} \bar{\Psi}^a  \left(   m \Gamma^{3456}  \right)^2
X_I^b \Gamma_I \epsilon. 
\nonu
\eea
Then the bosonic mass term $(m^2)_{IJ} \Gamma_J$
should take the form
$m^2 \sum_{I=3}^{10} \Gamma_I$. 
Then the diagonal bosonic mass term has nonzero
components for all eight elements. The
degeneracy $8$ is related to the ${\cal N}=8$ supersymmetry \cite{GSP,HLL,GRVV}.  
Then one obtains the bosonic mass term which 
appears in (\ref{bosonic})
\bea
(m^2)_{IJ} = \diag(m^2, m^2, m^2, m^2, m^2, m^2, m^2, m^2).
\label{mass8}
\eea

From the superpotential of the BL theory in ${\cal N}=2$ superspace with 
$SU(2) \times SU(2)=SO(4)$ gauge group, by adding the quadratic mass 
deformation (\ref{mass8}), we have  
the full
superpotential 
\footnote{As in previous case, one cannot conclude that the dimension of
  monomials in the superpotential is the sum of the dimensions of each
component. There is no conformal fixed point in the IR and the RG flow
does not terminate at a conformal field theory fixed point.}
\bea 
-\frac{1}{ 8 \cdot 4!} 
 \epsilon_{ABCD} \epsilon^{abcd} {\cal Z}_a^A {\cal Z}_b^B
{\cal Z}_c^C {\cal Z}_d^D 
-\frac{m^2}{2}  \sum_{A=1}^{4} ({\cal Z}_a^A)^2,
\label{massmass}
\eea
where ${\cal Z}_a^A $  is an ${\cal N}=2$ chiral 
superfield with $SU(4)$ index $A=1, 2, \cdots, 4$ and with $SO(4)$ gauge
group index $a=1,2, \cdots, 4$ as in previous section. 
The global symmetry $SO(8)_R$ of the theory
is broken to $[SO(4) \times SO(4) \times {\bf Z}_2]$ symmetry.
 The mass terms, the last terms in (\ref{massmass}), 
do not break ${\cal N}=8$.
The theory does not have matter multiplets transforming in the
adjoint of the gauge group because all of them are massive. 
The $SO(8)_R$ symmetry of the 
${\cal N}=8$ gauge theory is broken to 
$[SO(4) \times SO(4)]_R$ which is nothing but the R-symmetry.
We turn on the mass perturbation in the UV and flow to the
IR.
This maps to turning on the scalar field $\alpha$  in the $AdS_4$
supergravity where the scalar approaches to zero in the UV($r
\rightarrow \infty$) and develop a nontrivial profile as a function of
$r$ as one goes to the
IR($r \rightarrow -\infty$) through (\ref{flow1}).

From the superpotential of $U(2) \times U(2)$ Chern-Simons
matter theory, the quadratic mass deformations are
added as follows: 
\bea 
\frac{T^{-4}}{4!} 
 \epsilon_{ABCD} \Tr {\cal Z}^A {\cal Z}^{\ddagger B}
{\cal Z}^C {\cal Z}^{\ddagger D} 
-\frac{m^2}{2} T^{-2}  \sum_{A=1}^{4} \Tr {\cal Z}^A {\cal
  Z}^{\ddagger A}.
\label{massmass1}
\eea
Let us consider a vector $\vec{v}$ 
on which $SO(8)$ acts. Then the first $SO(4)$
acts on the first four components of the vector while the second 
$SO(4)$ acts on the last four components of the vector.
Here ${\bf Z}_2$ is the transformation which replaces the first four
components of the vector  with the last four components of the vector.
Then the $SU(4)$ subgroup above consists of complex rotation on the
four
component ``complex'' vector. Also there exists $U(1)$ subgroup 
which transform this complex vector as itself 
with an extra overall phase as in previous section.      
The independent two $SU(2)$ transformations, acting on 
the first two components of a complex vector and on 
the last two components of a complex vector respectively
and  the overall phase rotation give 
a manifest $SU(2) \times SU(2) \times U(1) \times {\bf Z}_2$ symmetry. 
Note that the subgroup of $SO(8)$ common to both $SU(4) \times U(1)$
for undeformed ${\cal N}=6$ Chern-Simons matter theory 
and $SO(4) \times SO(4)$ for the R-symmetry of the theory 
is nothing but $SU(2) \times SU(2) 
\times U(1)$.

After the integration over the superspace, the quadratic deformation 
leads to
\bea
&& m^2 T^{-2} \int d^2 \theta   \Tr  \sum_{A=1}^{4}
{\cal Z}^A {\cal Z}^{\ddagger A}  
\nonu \\
&&  =  - m^2 T^{-2} \Tr  \sum_{A=1}^{4}
\zeta^A \zeta^{\ddagger A}  + 2m^2 T^{-2} \Tr \sum_{A=1}^{4}  F^A
 Z^{\ddagger A} \nonu \\ 
&& = - m^2 T^{-2} \Tr  \sum_{A=1}^{4} \zeta^A \zeta^{\ddagger A} 
 + T^{-4} \frac{m^2 L}{3} \sum_{D=1}^{4} 
\epsilon^{ABC} \Tr \bar{Z}_A \bar{Z}_B^{\ddagger} \bar{Z}_C
Z^{\ddagger D}, 
\nonu
\eea
where the auxiliary field $F$ is replaced using the equation of motion
in the last line:$F^A = -\frac{L}{6} \epsilon^{ABCD} \bar{Z}_B
\bar{Z}_C^{\ddagger} 
\bar{Z}_D$ as in \cite{BKKS}. One sees that there are also quartic terms 
above. 

It would be interesting to study whether this mass-deformed theory can
be realized in terms of ${\cal N}=4$ superspace formalism explicitly. 

\section{The M2-brane probe }

It is known in \cite{CLP} that  the four-dimensional ${\cal N}=4$
$SO(4)$ gauged supergravity can be obtained from the 
eleven-dimensions.
The 11-dimensional metric is given by \cite{PW} 
\bea
ds_{11}^2 & = &
\Omega^2
\left( e^{2A(r)} dx_{\mu}^2+dr^2 \right)+  4 L^2 \Omega^2 d \theta^2
\nonu 
\\
&+&  \frac{L^2 \Omega^2}{ Y } \cos^2 \theta
\left( \si_1^2 + \si_2^2 +\si_3^2\right) +
 \frac{L^2 \Omega^2}{ \widetilde{Y}} \sin^2 \theta
\left( \widetilde{\si}_1^2 + \widetilde{\si}_2^2 +
\widetilde{\si}_3^2 \right).
\label{metric}
\eea
Here the left-invariant one-forms parametrize the three sphere
${\bf S}^3$ 
and similarly 
other left-invariant one-forms parametrize the other three sphere
${\bf S}^3$. 
These are invariant under the action of the $SO(4) \times SO(4)$
R-symmetry. An interchange symmetry 
$\theta \rightarrow \frac{\pi}{2}-\theta$ and $\alpha \rightarrow
-\alpha$ appears also.
The various functions appearing in this metric (\ref{metric}) are 
defined by 
\bea
\Omega & = & \left(Y \widetilde{Y} \right)^{\frac{1}{6}}, 
\nonu \\
Y & = & 
\cos^2 \theta \left[ \cosh (2\al) + \cos \zeta \sinh (2\al) \right] 
+ \sin^2 \theta,
\nonu \\
\widetilde{Y}  & = & 
\sin^2 \theta \left[ \cosh (2\al) - \cos \zeta \sinh (2\al)
\right] 
+\cos^2 \theta.
\label{func}
\eea

The three-form potential is given by 
\bea
A^{(3)} = \frac{k^3 Z}{\sinh^3 \al} d t \wedge d x_1 \wedge d x_2 +
L^3 \sin \zeta \sinh (2\al) \left( \frac{\cos^4 \theta}{Y} \si_1 \wedge
  \si_2 \wedge \si_3 -\frac{\sin^4 \theta}{\widetilde{Y}}
  \widetilde{\si}_1
\wedge \widetilde{\si}_2 \wedge \widetilde{\si}_3 \right),
\nonu
\eea
where the function is defined as
\bea
Z = 
\frac{1}{2 \cosh \al} \left( Y + \widetilde{Y} \right)=
\cosh \al + \cos \zeta \cos (2\theta) \sinh \al. 
\label{tildeZ}
\eea
When $\theta=\frac{\pi}{4}$, this $Z$ is identical to $W$ in (\ref{superpot1}).
As we did before, using the definitions (\ref{ABCD}), 
the corresponding 28-beins are obtained in (\ref{Res1}), (\ref{Res2}), 
(\ref{Res3}) and (\ref{Res4}) in
appendix D.

By computing the real part of $11$ component of the $\widetilde{A}_1$
tensor (\ref{Res5}), one obtains 
the geometrical superpotential $W_{gs}$ as follows:
\bea
W_{gs} & = & \cosh \alpha \left( Y_1^2 +
Y_2^2+Y_3^2+Y_4^2 +Y_5^2 +Y_6^2+Y_7^2+Y_8^2\right) 
\nonu \\
& +& \cos \zeta \sinh \alpha \left( -Y_1^2 -Y_2^2-Y_3^2-Y_4^2+
Y_5^2+Y_6^2+Y_7^2+Y_8^2 \right).
\label{geosuperpot1}
\eea
We want to see how
this geometric superpotential is related to the superpotential 
(\ref{superpot1}). One compares this 
(\ref{geosuperpot1}) with (\ref{tildeZ}) by looking at the $\alpha$ and
$\zeta$ dependence.
Through the relations
\bea
Y_1 & = & \sin \theta \cos (\frac{\alpha_1}{2})
\cos (\frac{\alpha_2 +\alpha_3}{2}), \qquad 
Y_2 =  \sin \theta \cos (\frac{\alpha_1}{2})
\sin (\frac{\alpha_2 +\alpha_3}{2}),
\nonu \\
Y_3 & = &  \sin \theta \sin (\frac{\alpha_1}{2})
\cos (\frac{\alpha_2 -\alpha_3}{2}), \qquad 
Y_4 =   \sin \theta \sin (\frac{\alpha_1}{2})
\sin (\frac{\alpha_2 -\alpha_3}{2}),
\nonu \\
Y_5 & = &  \cos \theta \cos (\frac{\beta_1}{2})
\cos (\frac{\beta_2 +\beta_3}{2}), \qquad 
Y_6 =   \cos \theta \cos (\frac{\beta_1}{2})
\sin (\frac{\beta_2 +\beta_3}{2}),
\nonu \\
Y_7 & = & \cos \theta \sin (\frac{\beta_1}{2})
\cos (\frac{\beta_2 -\beta_3}{2}), \qquad 
Y_8 =   \cos \theta \sin (\frac{\beta_1}{2})
\sin (\frac{\beta_2 -\beta_3}{2}),
\nonu 
\eea
this geometric superpotential (\ref{geosuperpot1}) 
reduces to 
\bea
W_{gs} =Z,
\nonu
\eea 
with (\ref{tildeZ}) and $\alpha_i$ and $\beta_i$ are Euler angles
that parametrize two independent sets of $SU(2)$ 
left-invariant one-forms 
respectively. 

The potential seen by the M2-brane probe 
has a factor $
e^{3A} (\Omega^3 - 2 Z)$ with (\ref{func}) and (\ref{tildeZ}).
The moduli spaces of the brane probe are given by the loci
where the potential vanishes. 
There is no solution for internal coordinates satisfying the vanishing
of potential. This is due to the fact that there are no massless
scalars in this maximal case. All of them are massive.

Furthermore, by calculating the ${\widetilde A}_2$ tensor obtained from the
geometric $\widetilde{T}$ tensor, one arrives at the geometric scalar potential 
\bea
V_{gp}(\al,\zeta,\theta) & = &
\frac{g^2}{4}  \left[ 7+ \cos (2\zeta) + 2 \cosh
  (4\al) \sin^2 \zeta - 4 \cosh (2\alpha) (-6 + \cos [4\theta] \sinh^2
  \alpha) 
\right. \nonu \\
& + & \left.  32\cos \zeta \cos (2\theta) \sinh (2\alpha) +2
  \cos(4\theta)
(6\sinh^2 \alpha + \cos[2\zeta] \sinh^2 [2\alpha])
\right].  
\nonu
\eea
This has a critical point at $\alpha= 0=\zeta$ and $\theta=\frac{\pi}{4}$.
When $\zeta=0$,
the transverse components of three-form potential 
vanish and $A^{(3)}$ due to the factor $\sin \zeta$ reduces to 
\bea
A^{(3)} = k^3 
H^{-1} d t \wedge d x \wedge d y, \qquad
H \equiv \frac{\sinh^3 \alpha}{ e^{-\alpha} \sin^2 \theta + e^{\alpha}
\cos^2 \theta}.
\nonu
\eea
Moreover, by the change of variables appropriately
the metric can be simplified as the standard harmonic form where the
M2-branes are spread out into a solid four-ball. This is consistent
with the case where $\chi=0$ for ${\cal N}=4$ supersymmetric case 
before. 

\section{Conclusions and outlook}


We have found the gauge duals in the context of BL theory
(\ref{fullexp}) and 
(\ref{massmass}) and
Chern-Simons matter theory (\ref{bl}) and (\ref{massmass1}) 
to the holographic ${\cal N}=4$
supersymmetric RG flow and the holographic ${\cal N}=8$
supersymmetric RG flow.

As pointed out in \cite{GW}, when the parameters satisfy 
$\phi=\pm 2\varphi$, then there exist four supersymmetries. In other
words, ${\cal N}=2$ supersymmetry.
In this case, the corresponding superpotential will be either 
$z_2$ or $z_3$ which has multiplicities 2, 2, 
respectively.
One can construct 11-dimensional solution by using the
formula given by \cite{dWNW}. 
According to \cite{GW}, there exists a family of solutions 
that interpolates between the pure metric flow with $\phi=\varphi=0$
and the flow with $\phi=0, \varphi=\frac{\pi}{2}$. 
How one can realize this 
explicitly?

There are further developments \cite{GNW} which 
generalizes the work of \cite{GW}, 
using the Killing spinors. So it would be interesting to study 
\cite{GNW} in the present context.
Moreover, further generalization of \cite{PW} arises in the work of 
\cite{BW}. It would be interesting to find out how the gauge dual appears.
The regularity of the solutions was proved in \cite{LLM} and it would
be interesting to find out the implication between the flux
configuration and gauge theory configuration.

Besides the two supersymmetric critical points ${\cal N}=2$ $SU(3)
\times U(1)$ invariant point, ${\cal N}=1$ $G_2$ invariant point of 
four-dimensional ${\cal N}=8$ gauged supergravity,
there exist also three nontrivial nonsupersymmetric 
critical points as well as the trivial ${\cal N}=8$ 
$SO(8)$ critical point for the scalar
potential:$SO(7)^{+}, SO(7)^{-}$ and $SU(4)^{-}$. 
It would be interesting to discover any flow equations connecting any
two (non)supersymmetric critical points.  

Recall that the set of seventy scalars in ${\cal N}=8$ gauged
supergravity
has six singlets of $SU(3)$. Three singlets from ${\bf 35}_v$ and
three singlets from ${\bf 35}_c$. Under the $SO(3)=SU(2)$ subgroup of 
$SU(3)$, the irreducible representation ${\bf 6}$ of $SU(3)$ breaks
into ${\bf 5}$ plus ${\bf 1}$.   Totally the $SO(3)$-singlet space has
four more fields and may be parametrized by ten fields. It would be
interesting to find out any new $AdS_4$ critical points, if any, in the
$SO(3)$-invariant sector of ${\cal N}=8$ gauged supergravity in 
four-dimensions and see how the gauge duals appear. 

\vspace{.7cm}

\centerline{\bf Acknowledgments}

We would like to thank K. Hosomichi  and H. Lin for communications.
This work was supported by grant No.
R01-2006-000-10965-0 from the Basic Research Program of the Korea
Science \& Engineering Foundation.  

\appendix

\renewcommand{\thesection}{\large \bf \mbox{Appendix~}\Alph{section}}
\renewcommand{\theequation}{\Alph{section}\mbox{.}\arabic{equation}}
\section{The $28 \times 28$ matrices $u$ and $v$, $A_2$ tensor,
  kinetic terms and $SU(8)$ connection 
with $[SU(2) \times U(1)]^2$ symmetry in $SU(8)$ basis}

The 28-beins $u^{IJ}_{\;\;\;KL}$ and $v^{IJKL}$ 
fields, which are elements of $56 \times 56$ ${\cal V}(x)$ of
the fundamental 56-dimensional representation of $E_{7(7)}$ through (\ref{56bein}),
can be obtained 
by exponentiating the vacuum expectation values $\phi_{ijkl}$ 
(\ref{phiijkl}) via (\ref{calV}). These 28-beins have the
following seven $4 \times 4$ block diagonal matrices $u_i$ and $v_i$
where $i=1, 2, \cdots, 7$ respectively:
\bea 
u^{IJ}_{\;\;\;KL} & = & \mbox{diag}
(u_1,u_2,u_3,u_4,u_5,u_6,u_7), \nonu \\
v^{IJKL} & = &
\mbox{diag} (v_1,v_2,v_3,v_4,v_5,v_6,v_7). \nonu 
\eea 
Each hermitian(for example, 
$(u_1)^{78}_{\;\;\;12}=((u_1)^{12}_{\;\;\;78})^{\ast}=
\frac{1}{2} e^{i(\phi+\varphi)}E$) 
submatrix is $4 \times 4$ matrix and we denote
antisymmetric index pairs $[IJ]$ and $[KL]$ explicitly for convenience.
For simplicity, we make an empty space corresponding to lower triangle 
elements that can be read off from the corresponding upper triangle
elements by hermiticity. Then the $4 \times 4$ submatrix leads to
\bea
&& u_{1}=\left(\begin{array}{ccccc}
 & [12] & [34] & [56] & [78]\\{}
[12] & A & B & \frac{1}{2} e^{-i(\phi-\varphi)} E & \frac{1}{2}
e^{-i(\phi+\varphi)} E\\{}
[34] &  & A & \frac{1}{2} e^{-i(\phi-\varphi)} E & \frac{1}{2}
e^{-i(\phi+\varphi)} E\\{}
[56] &  &  & A & e^{-2i\varphi}B\\{}
[78] &  & 
&  & A
\end{array}\right),
u_{2}=\left(\begin{array}{ccccc}
 & [13] & [24] & [57] & [68]\\
{}[13] & G & 0 & 0 & 0\\
{}[24] &  & G & 0 & 0\\
{}[57] &  &  & G & 0\\
{}[68] &  &  &  & G\end{array}\right),
\nonu \\
&& u_{4}=\left(\begin{array}{ccccc}
 & [15] & [26] & [37] & [48]\\{}
[15] & I & 0 & 0 & 0\\{}
[26] &  & I & 0 & 0\\{}
[37] &  &  & I & 0\\{}
[48] &  &  &  & I\end{array}\right)=
u_{5}= u_{6}=
u_{7}, \qquad u_3 =u_2,
\nonu 
\eea
where the group indices of $u_3$ are different from those of $u_2$
and so on and
\bea
&& v_{1}=\left(\begin{array}{ccccc}
 & [12] & [34] & [56] & [78]\\{}
[12] & e^{-i\phi}D & e^{-i\phi}C & \frac{1}{2} e^{-i\varphi} F &
\frac{1}{2}
e^{i\varphi} F\\{}
[34] &  & e^{-i\phi}D & \frac{1}{2} e^{-i\varphi} F & \frac{1}{2} 
e^{i\varphi} F\\{}
[56] &  &  & e^{i(\phi-2\varphi)}D & e^{i\phi}C\\{}
[78] &  &  &  & e^{i(\phi+2\varphi)}D
\end{array}\right), \nonu \\
&& v_{2}=\left(\begin{array}{ccccc}
 & [13] & [24] & [57] & [68]\\{}
[13] & -e^{-i\phi}H & 0 & 0 & 0\\{}
[24] &  & -e^{-i\phi}H & 0 & 0\\{}
[57] &  &  & -e^{i\phi}H & 0\\{}
[68] &  &  &  & -e^{i\phi}H\end{array}\right)=-v_3,
\nonu \\
&& v_{4}=\left(\begin{array}{ccccc}
 & [15] & [26] & [37] & [48]\\{}
[15] & -e^{-i\varphi}J & 0 & 0 & 0\\{}
[26] &  & -e^{-i\varphi}J & 0 & 0\\{}
[37] &  &  & -e^{i\varphi}J & 0\\{}
[48] &  &  &  & -e^{i\varphi}J\end{array}\right)=-v_{5},
\nonu \\
&& v_{6}=\left(\begin{array}{ccccc}
 & [17] & [28] & [35] & [46]\\
{}[17] & -e^{i\varphi}J & 0 & 0 & 0 \\
{}[28] &  & -e^{i\varphi}J & 0 & 0 \\
{}[35] &  &  & -e^{-i\varphi}J & 0 \\
{}[46] &  &  &  & -e^{-i\varphi}J
\end{array}\right)=-v_{7},
\label{uv}
\eea
where
we introduce some quantities that are functions of $\al$ and $\chi$
as follows:
\bea
A & \equiv & \cosh \alpha \cosh^{2} \chi,\qquad 
B \equiv \cosh \alpha \sinh^{2} \chi,\qquad
C \equiv \sinh \alpha \cosh^{2} \chi,
\nonu \\
D & \equiv & \sinh \alpha \sinh^{2} \chi,\qquad 
E \equiv \sinh \alpha \sinh (2\chi),\qquad 
F \equiv \cosh \alpha \sinh (2\chi), \nonu \\
G & \equiv & \cosh \alpha,\qquad
H \equiv \sinh \alpha,\qquad 
I \equiv \cosh \chi,\qquad
J  \equiv  \sinh \chi.
\nonu
\eea
The lower triangle part can be read off from the upper triangle part by
hermitian property.
Also, the other kinds of 
28-beins $u_{IJ}^{\;\;\;KL}$ and $v_{IJKL}$ are obtained by taking a
complex conjugation of (\ref{uv}). The complex conjugation operation 
can be done by raising or lowering the indices:  
$(u_{IJ}^{\;\;\;KL})^{\ast} =  u^{IJ}_{\;\;\;KL}$ and so on.

The components of $A_2$ tensor, $A_{2, L}^{\;\;\;\;\;IJK}$, 
are obtained from the defining equation
\bea
T_l^{\;kij} & = & 
\left(u^{ij}_{\;\;IJ} +v^{ijIJ} \right) \left( u_{lm}^{\;\;\;JK} 
u^{km}_{\;\;\;KI}-v_{lmJK} v^{kmKL} \right),
\label{ttens}
\eea
and the well-known tensors of the ${\cal N}=8$ gauged supergravity are
\bea
A_1^{\;\;ij} & =& 
-\frac{4}{21} T_{m}^{\;\;ijm}, \;\;\; A_{2l}^{\;\;\;ijk}=-\frac{4}{3}
T_{l}^{\;[ijk]},
\label{a1a2}
\eea
by substituting (\ref{uv}) into (\ref{ttens}) and (\ref{a1a2})
and 
they are classified by seven different fields with degeneracies 4, 4, 2,
2, 4, 4, 4 respectively  and  
given by:
\bea
A_{2,\;1}^{\;\;\;278} & = & 
A_{2,\;2}^{\;\;\;187}=A_{2,\;3}^{\;\;\;478}=A_{2,\;4}^{\;\;\;387}\equiv
y_{1} = -\frac{1}{2}e^{i\varphi}\frac{\partial z_{1}^{*}}{\partial\chi}, 
\nonu \\
A_{2,\;1}^{\;\;\;234}& = & 
A_{2,\;2}^{\;\;\;143}=A_{2,\;3}^{\;\;\;124}=A_{2,\;4}^{\;\;\;132}\equiv
y_{2} = -e^{-i\phi}\frac{\partial z_{1}^{*}}{\partial\alpha},\nonu \\
A_{2,\;7}^{\;\;\;568} & = & A_{2,\;8}^{\;\;\;576}\equiv y_{3} =
-e^{i\phi}\frac{\partial z_{3}^{\ast}}{\partial\alpha}, \nonu \\
A_{2,\;5}^{\;\;\;678} & = & A_{2,\;6}^{\;\;\;587}\equiv y_{4} =
-e^{i\phi}\frac{\partial z_{2}^{\ast}}{\partial\alpha}, \nonu \\
A_{2,\;7}^{\;\;\;128}& = & 
A_{2,\;7}^{\;\;\;348}=A_{2,\;8}^{\;\;\;172}=A_{2,\;8}^{\;\;\;374}\equiv
y_{5} = -\frac{1}{2}e^{i\varphi}\frac{\partial z_{3}^{\ast}}{\partial\chi}, \nonu \\
A_{2,\;1}^{\;\;\;256}& = & 
A_{2,\;2}^{\;\;\;165}=A_{2,\;3}^{\;\;\;456}=A_{2,\;4}^{\;\;\;365}\equiv
y_{6} = -\frac{1}{2}e^{-i\varphi}\frac{\partial z_{1}^{*}}{\partial\chi}, \nonu \\
A_{2,\;5}^{\;\;\;126}& = & 
A_{2,\;5}^{\;\;\;346}=A_{2,\;6}^{\;\;\;152}=A_{2,\;6}^{\;\;\;354}\equiv
y_{7} = -\frac{1}{2}e^{-i\varphi}\frac{\partial z_{2}^{*}}{\partial\chi},
\label{A2}
\eea
where the redefined functions are given by
\bea
y_{1} & = & 
-e^{i\varphi}\cosh \chi
\left(\cosh \alpha +e^{i\phi}
\sinh \alpha \right) \sinh \chi,
\nonu \\
y_{2} & = & -e^{-i\phi}\cosh^{2} \chi
\sinh \alpha-\cosh \alpha \sinh^{2} \chi,\nonu \\
y_{3} & = & -e^{i\phi} \cosh^{2} \chi
\sinh \alpha-e^{-2i\varphi}
\cosh \alpha \sinh^{2} \chi,\nonu \\
y_{4} & = &
-e^{i\phi}\cosh^{2} \chi \sinh \alpha-
e^{2i\varphi}\cosh \alpha \sinh^{2} \chi,\nonu \\
y_{5} & = & -\frac{1}{2}e^{-i(\phi+\varphi)}\left(e^{i(\phi+2\varphi)}
\cosh \alpha+\sinh \alpha \right)\sinh\left(2\chi\right),
\nonu \\
y_{6} & = & -e^{-i\varphi}\cosh \chi
\left(\cosh \alpha+ e^{i\phi}\sinh \alpha \right)
\sinh \chi, \nonu \\
y_{7} & = & -\frac{1}{2}e^{-i(\phi+\varphi)}\left(e^{i\phi}\cosh
\alpha+e^{2i\varphi} \sinh \alpha \right)\sinh\left(2\chi\right).
\label{y7}
\eea
We write these in terms of the derivatives of the eigenvalues of $A_1$
tensor (\ref{zis}) with respect to the $\alpha$ and $\chi$.
It is manifest that
$A_{2,L}^{\;\;\;\;IJK}=-A_{2,L}^{\;\;\;\;IKJ}$, by definition.
Moreover there exists a symmetry between the upper indices:
$A_{2,L}^{\;\;\;\;IJK}=A_{2,L}^{\;\;\;\;JKI}=A_{2,L}^{\;\;\;\;KIJ}$.
Recall that in the supersymmetric transformation rules, the
$A_{2, L}^{\;\;\;\;\;IJK}$ appears in the equation satisfied by 
spin-$\frac{1}{2}$ field (\ref{susy}). This is the reason 
why we rewrite it in terms of superpotential.

The kinetic terms coming from the definition 
\bea
A_{\mu}^{\;\; ijkl} = -2 \sqrt{2} \left( u^{ij}_{\;\;IJ} 
\partial_{\mu} v^{klIJ} -
v^{ijIJ} \partial_{\mu} u^{kl}_{\;\;IJ} \right),
\nonu
\eea
can be summarized as following seven 
$4\times 4$ block diagonal hermitian matrices like as 28-beins 
 $u^{IJ}_{\;\;\;KL}$ and $v^{IJKL}$ as above:
\bea
A_{\mu}^{\; IJKL}=
\mbox{diag} \left(A_{\mu,1},A_{\mu,2},
A_{\mu,3},A_{\mu,4},A_{\mu,5},A_{\mu,6},A_{\mu,7}\right),
\nonu
\eea
where each hermitian submatrix can be written as follows:
\bea
A_{\mu,1} & = & \left(\begin{array}{ccccc}
 & [12] & [34] & [56] & [78]\\{}
[12] & 0 & -a^{*} & -b^{*} & -b\\{}
[34] &  & 0 & -b^{*} & -b\\{}
[56] &  &  & 0 & -a\\{}
[78] &  &  &  & 0\end{array}\right), \nonu \\
A_{\mu,2} & = & \left(\begin{array}{ccccc}
 & [13] & [24] & [57] & [68]\\
{}[13] & 0 & a^{*} & 0 & 0\\
{}[24] &  & 0 & 0 & 0\\
{}[57] &  &  & 0 & a\\
{}[68] &  &  &  & 0\end{array}\right)=-A_{\mu,3}, 
\nonu \\
A_{\mu,4} & = & \left(\begin{array}{ccccc}
 & [15] & [26] & [37] & [48]\\{}
[15] & 0 & b^{*} & 0 & 0\\{}
[26] &  & 0 & 0 & 0\\{}
[37] &  &  & 0 & b\\{}
[48] &  &  &  & 0\end{array}\right)=-A_{\mu,5}=(A_{\mu,6})^{\ast} 
=(-A_{\mu,7})^{\ast},
\label{amu}
\eea
where we introduce
\bea
a \equiv \sqrt{2}e^{i\phi} \left(
  \partial_{\mu}\alpha +\frac{i}{2}
\sinh\left(2\alpha\right) \partial_{\mu} \phi \right) ,\qquad 
b \equiv \sqrt{2}e^{i\varphi} \left( \partial_{\mu}\chi+
\frac{i}{2}\sinh\left(2\chi\right) \partial_{\mu}\varphi \right).
\nonu
\eea
The lower triangle part of (\ref{amu}) 
can be read off from the upper triangle part by
hermitian property.
Then the final expression of kinetic terms by counting the correct
multiplicities is given by
\bea
\frac{1}{96}|A_{\mu}^{\; ijkl}|^{2}=
\left(\partial_{\mu}\alpha\right)^{2}+\frac{1}{4}\sinh^{2}
\left(2\alpha\right)\left(\partial_{\mu}\phi\right)^{2}+
2\left[ \left(\partial_{\mu}\chi\right)^{2}+
\frac{1}{4}\sinh^{2}\left(2\chi\right)\left(\partial_{\mu}\varphi\right)^{2}
\right]. 
\nonu
\eea
In particular, when $\phi=0$ and $\varphi=\frac{\pi}{2}$, this becomes 
$\left(\partial_{\mu}\alpha\right)^{2} + 2
\left(\partial_{\mu}\chi\right)^{2}$
as in (\ref{kineticterm}).

The quantity in the covariant derivative (\ref{cov}) was defined as
\bea
{\cal B}_{\mu\;\;\;j}^{\;\;i} \equiv \frac{2}{3} \left( u^{ik}_{\;\;\;IJ}
\partial_{\mu} u_{jk}^{\;\;\;IJ} - v^{ikIJ} \partial_{\mu} v_{jkIJ} \right),
\label{calcalB}
\eea
and it turns out, by substituting the 28-beins (\ref{uv}) into (\ref{calcalB}), 
that it is diagonal and is given by
\bea
{\cal B}_{\mu\;\;\;j}^{\;\;i} & = & \mbox{diag}(
-i\partial_{\mu}\phi \sinh^{2} \alpha,
 -i\partial_{\mu}\phi \sinh^{2} \alpha, 
-i\partial_{\mu}\phi \sinh^{2} \alpha,
-i\partial_{\mu}\phi \sinh^{2} \alpha, 
\nonu \\
&& i\left(
\partial_{\mu}\phi \sinh^{2} \alpha-
2 \partial_{\mu}\varphi \sinh^{2} \chi \right),
i\left(
\partial_{\mu}\phi \sinh^{2} \alpha-
2 \partial_{\mu}\varphi \sinh^{2} \chi \right), \nonu \\
&& i\left(\partial_{\mu}\phi
\sinh^{2} \alpha+2\partial_{\mu}
\varphi \sinh^{2} \chi \right),
i\left(\partial_{\mu}\phi
\sinh^{2} \alpha+2\partial_{\mu}
\varphi \sinh^{2} \chi \right)).
\label{calB}
\eea

\section{The $28 \times 28$ matrices $U$ and $V$ and $A_1$ tensor
with $[SU(2) \times U(1)]^2$ symmetry in $SL(8,{\bf R})$ basis}

The 28-beins $A_{IJKL},
B_{IJ}^{\;\;\;\; KL},
C_{\;\;\;\;KL}^{IJ}$ and $
D^{IJKL}$ 
fields (\ref{ABCD}), that can be obtained from the $SU(8)$ basis to the
$SL(8,{\bf R})$ basis using gamma matrices, have the
following four kinds of 
seven $4 \times 4$ block diagonal matrices $a_i, b_i, c_i$ and $d_i$
where $i=1, 2, \cdots, 7$ respectively:
\bea 
A_{IJKL} & = & \mbox{diag}
(a_1,a_2,a_3,a_4,a_5,a_6,a_7), \qquad
B_{IJ}^{\;\;\;\; KL} = 
\mbox{diag} (b_1,b_2,b_3,b_4,b_5,b_6,b_7), \nonu \\
C_{\;\;\;\;KL}^{IJ} & = & 
\mbox{diag} (c_1,c_2,c_3,c_4,c_5,c_6,c_7), \qquad
D^{IJKL}  =  
\mbox{diag} (d_1,d_2,d_3,d_4,d_5,d_6,d_7). 
\label{abcd}
\eea 
It turns out that the 28-bein $A_{IJKL}$  with (\ref{abcd}) are given by
\bea
a_2 & = & \left(\begin{array}{ccccc}
 & [13] & [24] & [57] & [68]\\
{}[13] & a & -a & a & -a\\
{}[24] & -b & b & -b & b\\
{}[57] & -a & -a & -a & -a\\
{}[68] & b^{*} & b^{*} & b^{*} & b^{*}\end{array}\right), \qquad
a_4=\left(\begin{array}{ccccc}
 & [15] & [26] & [37] & [48]\\
{}[15] & -c & -d & d & c\\
{}[26] & c & -d & d & -c\\
{}[37] & -c^{*} & d^{*} & d^{*} & -c^{*}\\
{}[48] & c^{*} & d^{*} & d^{*} & c^{*}\end{array}\right),
\nonu \\
a_6 & = & \left(\begin{array}{ccccc}
 & [17] & [28] & [35] & [46]\\
{}[17] & -f & -f & -f & -f\\
{}[28] & g^{*} & g^{*} & g^{*} & g^{*}\\
{}[35] & f & -f & f & -f\\
{}[46] & -g & g & -g & g\end{array}\right), \qquad
a_1  =  0= a_3=a_5=a_7,
\label{A}
\eea
where we introduce
\bea
a & \equiv & \sqrt{2}\cosh \alpha,\qquad
b \equiv \sqrt{2}e^{i\phi}\sinh \alpha, \qquad
c  \equiv  \sqrt{2}\left(\cosh \chi+e^{i\varphi}\sinh \chi \right),
\nonu \\
d & \equiv & \sqrt{2}\left(\cosh \chi -e^{i\varphi}\sinh \chi \right),
\qquad
f  \equiv  \sqrt{2}\cosh \chi,\qquad 
g \equiv \sqrt{2}e^{i\varphi}\sinh \chi.
\label{redef}
\eea
Compared with the previous 28-beins, there is no hermiticity property
in $SL(8, {\bf R})$ basis.
Similarly, one gets the 28-beins $B_{IJ}^{\;\;\;\; KL}$  with
(\ref{abcd}) and (\ref{redef})
\bea
b_2 & = & \left(\begin{array}{ccccc}
 & [13] & [24] & [57] & [68]\\
{}[13] & a & -a & a & -a\\
{}[24] & b & -b & a & -b\\
{}[57] & -a & -a & -a & -a\\
{}[68] & -b^{*} & -b^{*} & -b^{*} & -b^{*}\end{array}\right),
\qquad
b_4=\left(\begin{array}{ccccc}
 & [15] & [26] & [37] & [48]\\
{}[15] & -d & -c & c & d\\
{}[26] & d & -c & c & -d\\
{}[37] & -d^{*} & c^{*} & c^{*} & -d^{*}\\
{}[48] & d^{*} & c^{*} & c^{*} & d^{*}\end{array}\right),\nonu \\
b_6 & = & \left(\begin{array}{ccccc}
 & [17] & [28] & [35] & [46]\\
{}[17] & -f & -f & -f & -f\\
{}[28] & -g^{*} & -g^{*} & -g^{*} & -g^{*}\\
{}[35] & f & -f & f & -f\\
{}[46] & g & -g & g & -g\end{array}\right), 
\qquad
b_1  =  0=b_3 =b_5=b_7.
\label{B}
\eea
Moreover, one has the 28-beins $C_{\;\;\;\;KL}^{IJ}$  with
(\ref{abcd})
and (\ref{redef})
\bea
c_2 & = & \left(\begin{array}{ccccc}
 & [13] & [24] & [57] & [68]\\
{}[13] & a & -a & a & -a\\
{}[24] & -b^{*} & b^{*} & -b^{*} & b^{*}\\
{}[57] & -a & -a & -a & -a\\
{}[68] & b & b & b & b\end{array}\right), \qquad
c_4=\left(\begin{array}{ccccc}
 & [15] & [26] & [37] & [48]\\
{}[15] & -c^{*} & -d^{*} & d^{*} & c^{*}\\
{}[26] & c^{*} & -d^{*} & d^{*} & -c^{*}\\
{}[37] & -c & d & d & -c\\
{}[48] & c & d & d & c\end{array}\right),\nonu \\
c_6 & = & \left(\begin{array}{ccccc}
 & [17] & [28] & [35] & [46]\\
{}[17] & -f & -f & -f & -f\\
{}[28] & g & g & g & g\\
{}[35] & f & -f & f & -f\\
{}[46] & -g^{*} & g^{*} & -g^{*} & g^{*}\end{array}\right), \qquad
c_1  =  0 = c_3 =c_5=c_7.
\label{C}
\eea
Finally, we have the 28-beins $D^{IJKL}$ with (\ref{abcd}) 
and (\ref{redef})
\bea
d_2 & = & \left(\begin{array}{ccccc}
 & [13] & [24] & [57] & [68]\\
{}[13] & -a & a & -a & a\\
{}[24] & -b^{*} & b^{*} & -b^{*} & b^{*}\\
{}[57] & a & a & a & a\\
{}[68] & b & b & b & b\end{array}\right),\qquad
d_4=\left(\begin{array}{ccccc}
 & [15] & [26] & [37] & [48]\\
{}[15] & d^{*} & c^{*} & -c^{*} & -d^{*}\\
{}[26] & -d^{*} & c^{*} & -c^{*} & d^{*}\\
{}[37] & d & -c & -c & d\\
{}[48] & -d & -c & -c & -d\end{array}\right),\nonu \\
d_6 & = & \left(\begin{array}{ccccc}
 & [17] & [28] & [35] & [46]\\
{}[17] & f & f & f & f\\
{}[28] & g & g & g & g\\
{}[35] & -f & f & -f & f\\
{}[46] & -g^{*} & g^{*} & -g^{*} & g^{*}\end{array}\right),
\qquad 
d_1  =  0 = d_3 = d_5 =d_7.
\label{D}
\eea

From the definition of (\ref{ttilde}), one obtains the components of 
$\widetilde{A}_1$ tensor by substituting (\ref{A}), (\ref{redef}), (\ref{B}), (\ref{C}) 
and (\ref{D}) into (\ref{ttilde}) 
and the diagonal components are 
\bea
\widetilde{A}_{1}^{\;11} & = & \widetilde{A}_{1}^{\;22}=2e^{-i\phi}
(\sinh\alpha(-Y_{1}^{2}-Y_{2}^{2}-Y_{3}^{2}-Y_{4}^{2}+
\cosh(2\chi)(Y_{5}^{2}+Y_{6}^{2}+Y_{7}^{2}+Y_{8}^{2}) \nonu \\
 &  &
 +\sinh(2\chi)(-i\sin\varphi(Y_{1}^{2}+Y_{2}^{2}-Y_{3}^{2}-Y_{4}^{2})+
\cos\varphi(-Y_{5}^{2}-Y_{6}^{2}+Y_{7}^{2}+Y_{8}^{2}))) \nonu \\
 &  & +e^{i\phi}\cosh\alpha(Y_{1}^{2}+Y_{2}^{2}+Y_{3}^{2}+Y_{4}^{2}+
\cosh(2\chi)(Y_{5}^{2}+Y_{6}^{2}+Y_{7}^{2}+Y_{8}^{2}) \nonu \\
 &  &
 +\sinh(2\chi)(i\sin\varphi(Y_{1}^{2}+Y_{2}^{2}-Y_{3}^{2}-Y_{4}^{2})+
\cos\varphi(-Y_{5}^{2}-Y_{6}^{2}+Y_{7}^{2}+Y_{8}^{2})))),
\nonu \\
\widetilde{A}_{1}^{\;33} & = & \widetilde{A}_{1}^{\;44}=-2e^{-i\phi}
(\sinh\alpha(Y_{1}^{2}+Y_{2}^{2}+Y_{3}^{2}+Y_{4}^{2}-\cosh(2\chi)(Y_{5}^{2}+
Y_{6}^{2}+Y_{7}^{2}+Y_{8}^{2}) \nonu \\
 &  &
 +\sinh(2\chi)(-i\sin\varphi(Y_{1}^{2}+Y_{2}^{2}-Y_{3}^{2}-Y_{4}^{2})+
\cos\varphi(Y_{5}^{2}+Y_{6}^{2}-Y_{7}^{2}-Y_{8}^{2}))) \nonu \\
 &  & +e^{i\phi}\cosh\alpha(-Y_{1}^{2}-Y_{2}^{2}-Y_{3}^{2}-Y_{4}^{2}-
\cosh(2\chi)(Y_{5}^{2}+Y_{6}^{2}+Y_{7}^{2}+Y_{8}^{2})\nonu \\
 &  &
 +\sinh(2\chi)(i\sin\varphi(Y_{1}^{2}+Y_{2}^{2}-Y_{3}^{2}-Y_{4}^{2})+
\cos\varphi(Y_{5}^{2}+Y_{6}^{2}-Y_{7}^{2}-Y_{8}^{2})))),
\label{tilone} 
\eea
where $\widetilde{A}_{1}^{\;11}$ provides the geometric 
superpotential (\ref{geosuperpot}) when we restrict to the case 
$\phi=0$ and $\varphi=\frac{\pi}{2}$ 
and moreover there are
\bea
\widetilde{A}_{1}^{\;55} & = & \widetilde{A}_{1}^{\;66}=e^{-2i\varphi}
(2\sinh\alpha(-e^{i\varphi}\sinh(2\chi)(Y_{5}^{2}+Y_{6}^{2}-
Y_{7}^{2}-Y_{8}^{2}) \nonu \\
 &  &
 +\sinh^{2}\chi(-Y_{1}^{2}-Y_{2}^{2}-Y_{3}^{2}-Y_{4}^{2}+Y_{5}^{2}+
Y_{6}^{2}+Y_{7}^{2}+Y_{8}^{2}) \nonu \\
 &  &
 +e^{2i\varphi}\cosh^{2}\chi(Y_{1}^{2}+Y_{2}^{2}+Y_{3}^{2}+Y_{4}^{2}+
Y_{5}^{2}+Y_{6}^{2}+Y_{7}^{2}+Y_{8}^{2})) \nonu \\
 &  &
 +2e^{i\phi}\sinh\alpha(-2e^{2i\varphi}\cosh^{2}\chi(Y_{1}^{2}+Y_{2}^{2}+
Y_{3}^{2}+Y_{4}^{2}-Y_{5}^{2}-Y_{6}^{2}-Y_{7}^{2}-Y_{8}^{2}) \nonu \\
 &  &
 -e^{i\varphi}\sinh(2\chi)(Y_{5}^{2}+Y_{6}^{2}-Y_{7}^{2}-Y_{8}^{2})
 \nonu \\ &  & + 
\sinh^{2}\chi(Y_{1}^{2}+Y_{2}^{2}+Y_{3}^{2}+Y_{4}^{2}+Y_{5}^{2}+Y_{6}^{2}+
Y_{7}^{2}+Y_{8}^{2}))),
\nonu \\
\widetilde{A}_{1}^{\;77} & = &
\widetilde{A}_{1}^{\;88}=-2e^{-i\varphi}
(\cosh\alpha+e^{i\phi}\sinh\alpha)\sinh(2\chi)(Y_{5}^{2}+
Y_{6}^{2}-Y_{7}^{2}-Y_{8}^{2}) \nonu \\
 &  &
 +2\cosh^{2}\chi(-e^{i\phi}\sinh\alpha(Y_{1}^{2}+Y_{2}^{2}+Y_{3}^{2}+
Y_{4}^{2}-Y_{5}^{2}-Y_{6}^{2}-Y_{7}^{2}-Y_{8}^{2})\nonu \\
 &  & +\cosh\alpha(Y_{1}^{2}+Y_{2}^{2}+Y_{3}^{2}+Y_{4}^{2}+Y_{5}^{2}+
Y_{6}^{2}+Y_{7}^{2}+Y_{8}^{2}))\nonu \\
 &  & +2e^{2i\varphi}\sinh^{2}\chi(\cosh\alpha(-Y_{1}^{2}-Y_{2}^{2}-
Y_{3}^{2}-Y_{4}^{2}+Y_{5}^{2}+Y_{6}^{2}+Y_{7}^{2}+Y_{8}^{2})\nonu \\
 &  & +e^{i\phi}\sinh\alpha(Y_{1}^{2}+Y_{2}^{2}+Y_{3}^{2}+Y_{4}^{2}+
Y_{5}^{2}+Y_{6}^{2}+Y_{7}^{2}+Y_{8}^{2}).
\label{tiltwo}
\eea
The off-diagonal components are given by
\bea
\widetilde{A}_{1}^{\;12} & = & 
\widetilde{A}_{1}^{\;34}
=
\widetilde{A}_{1}^{\;56}=
\widetilde{A}_{1}^{\;78}
=0, \nonu \\
\widetilde{A}_{1}^{\;13} & = & 
\widetilde{A}_{1}^{\;24}=2e^{-i(\phi+\varphi)}(-1+e^{2i\varphi})
(e^{i\phi}\cosh\alpha-\sinh\alpha)\sinh(2\chi)(Y_{1}Y_{3}+Y_{2}Y_{4}),
\nonu \\
\widetilde{A}_{1}^{\;14} & = & 
\widetilde{A}_{1}^{\;23}=2e^{-i(\phi+\varphi)}(-1+e^{2i\varphi})
(e^{i\phi}\cosh\alpha-\sinh\alpha)\sinh(2\chi)(Y_{1}Y_{4}-Y_{2}Y_{3}),
\nonu \\
\widetilde{A}_{1}^{\;58} & = & 
\widetilde{A}_{1}^{\;67}=2e^{-i\varphi}(-1+e^{2i\varphi})
(\cosh\alpha+e^{i\phi}\sinh\alpha)\sinh(2\chi)(Y_{5}Y_{8}+Y_{6}Y_{7}),
\nonu \\
\widetilde{A}_{1}^{\;57} & = & 
-\widetilde{A}_{1}^{\;68}=2e^{-i\varphi}(-1+e^{2i\varphi})
(\cosh\alpha+e^{i\phi}\sinh\alpha)\sinh(2\chi)(Y_{5}Y_{7}-Y_{6}Y_{8}),
\label{tilthree}
\eea
and moreover we have 
\bea
\widetilde{A}_{1}^{\;15}
& = & 2i(\cosh\chi-e^{-i\varphi}\sinh\chi)(\sin\phi\sinh(2\alpha)+
\sin\varphi\sinh(2\chi))(Y_{1}Y_{5}+Y_{2}Y_{6}) \nonu \\
 &  &
 -2i(\cosh\chi+e^{-i\varphi}\sinh\chi)(\sin\phi\sinh(2\alpha)-
\sin\varphi\sinh(2\chi))(Y_{3}Y_{7}+Y_{4}Y_{8}),
\nonu \\
\widetilde{A}_{1}^{\;26}
& = & 2i(\cosh\chi-e^{-i\varphi}\sinh\chi)(\sin\phi\sinh(2\alpha)+
\sin\varphi\sinh(2\chi))(Y_{1}Y_{5}+Y_{2}Y_{6}) \nonu \\
 &  &
 +2i(\cosh\chi+e^{-i\varphi}\sinh\chi)(\sin\phi\sinh(2\alpha)-
\sin\varphi\sinh(2\chi))(Y_{3}Y_{7}+Y_{4}Y_{8}),
\nonu \\
\widetilde{A}_{1}^{\;37}
& = & -2i(\cosh\chi-e^{i\varphi}\sinh\chi)(\sin\phi\sinh(2\alpha)-
\sin\varphi\sinh(2\chi))(Y_{1}Y_{5}-Y_{2}Y_{6})\nonu \\
 &  &
 +2i(\cosh\chi+e^{i\varphi}\sinh\chi)(\sin\phi\sinh(2\alpha)+
\sin\varphi\sinh(2\chi))(Y_{3}Y_{7}-Y_{4}Y_{8}),
\nonu \\
\widetilde{A}_{1}^{\;48}  
& = & -2i(\cosh\chi-e^{i\varphi}\sinh\chi)(\sin\phi\sinh(2\alpha)-
\sin\varphi\sinh(2\chi))(Y_{1}Y_{5}-Y_{2}Y_{6})\nonu \\
 &  &
 -2i(\cosh\chi+e^{i\varphi}\sinh\chi)(\sin\phi\sinh(2\alpha)+
\sin\varphi\sinh(2\chi))(Y_{3}Y_{7}-Y_{4}Y_{8}),
\label{tilfour}
\eea
and 
\bea
\widetilde{A}_{1}^{\;16} 
& = & 2i(\cosh\chi-e^{-i\varphi}\sinh\chi)(\sin\phi\sinh(2\alpha)+
\sin\varphi\sinh(2\chi))(Y_{1}Y_{6}-Y_{2}Y_{5})\nonu \\
 &  &
 -2i(\cosh\chi+e^{-i\varphi}\sinh\chi)(\sin\phi\sinh(2\alpha)-
\sin\varphi\sinh(2\chi))(Y_{3}Y_{8}-Y_{4}Y_{7}),
\nonu \\
\widetilde{A}_{1}^{\;25} 
& = & -2i(\cosh\chi-e^{-i\varphi}\sinh\chi)(\sin\phi\sinh(2\alpha)+
\sin\varphi\sinh(2\chi))(Y_{1}Y_{6}-Y_{2}Y_{5})\nonu \\
 &  &
 -2i(\cosh\chi+e^{-i\varphi}\sinh\chi)(\sin\phi\sinh(2\alpha)-
\sin\varphi\sinh(2\chi))(Y_{3}Y_{8}-Y_{4}Y_{7}),
\nonu \\
\widetilde{A}_{1}^{\;38} 
&= & -2i(\cosh\chi-e^{i\varphi}\sinh\chi)(\sin\phi\sinh(2\alpha)-
\sin\varphi\sinh(2\chi))(Y_{1}Y_{6}+Y_{2}Y_{5})\nonu \\
 &  & +2i(\cosh\chi+e^{i\varphi}\sinh\chi)(\sin\phi\sinh(2\alpha)+
\sin\varphi\sinh(2\chi))(Y_{3}Y_{8}+Y_{4}Y_{7}),
\nonu \\
\widetilde{A}_{1}^{\;47} 
&= & 
2i(\cosh\chi-e^{i\varphi}\sinh\chi)(\sin\phi\sinh(2\alpha)-\sin\varphi\sinh(2\chi))
(Y_{1}Y_{6}+Y_{2}Y_{5})\nonu \\
 &  & +2i(\cosh\chi+e^{i\varphi}\sinh\chi)
(\sin\phi\sinh(2\alpha)+\sin\varphi\sinh(2\chi))(Y_{3}Y_{8}+Y_{4}Y_{7}),
\nonu \\
\widetilde{A}_{1}^{\;17} 
&= & 2i(\cosh\chi+e^{i\varphi}\sinh\chi)
(\sin\phi\sinh(2\alpha)+\sin\varphi\sinh(2\chi))(Y_{1}Y_{7}-
Y_{2}Y_{8}) \nonu \\
 &  & +2i(\cosh\chi-e^{i\varphi}\sinh\chi)(\sin\phi\sinh(2\alpha)-
\sin\varphi\sinh(2\chi))(Y_{3}Y_{5}-Y_{4}Y_{6}),
\nonu \\
\widetilde{A}_{1}^{\;28} 
&= & -2i(\cosh\chi+e^{i\varphi}\sinh\chi)(\sin\phi\sinh(2\alpha)+
\sin\varphi\sinh(2\chi))(Y_{1}Y_{7}-Y_{2}Y_{8})\nonu \\
 &  & +2i(\cosh\chi-e^{i\varphi}\sinh\chi)
(\sin\phi\sinh(2\alpha)-\sin\varphi\sinh(2\chi))(Y_{3}Y_{5}-Y_{4}Y_{6}),
\nonu \\
\widetilde{A}_{1}^{\;35} 
&= & 2i(\cosh\chi+e^{-i\varphi}\sinh\chi)(\sin\phi\sinh(2\alpha)-
\sin\varphi\sinh(2\chi))(Y_{1}Y_{7}+Y_{2}Y_{8})\nonu \\
 &  & +2i(\cosh\chi-e^{-i\varphi}\sinh\chi)(\sin\phi\sinh(2\alpha)+
\sin\varphi\sinh(2\chi))(Y_{3}Y_{5}+Y_{4}Y_{6}),
\nonu \\
\widetilde{A}_{1}^{\;46} 
&= & -2i(\cosh\chi+e^{-i\varphi}\sinh\chi)(\sin\phi\sinh(2\alpha)-
\sin\varphi\sinh(2\chi))(Y_{1}Y_{7}+Y_{2}Y_{8})\nonu \\
 &  & +2i(\cosh\chi-e^{-i\varphi}\sinh\chi)
(\sin\phi\sinh(2\alpha)+\sin\varphi\sinh(2\chi))(Y_{3}Y_{5}+Y_{4}Y_{6}),
\nonu \\
\widetilde{A}_{1}^{\;18} 
&= & 2i(\cosh\chi+e^{i\varphi}\sinh\chi)
(\sin\phi\sinh(2\alpha)+\sin\varphi\sinh(2\chi))(Y_{1}Y_{8}+
Y_{2}Y_{7})\nonu \\
 &  & +2i(\cosh\chi-e^{i\varphi}\sinh\chi)(\sin\phi\sinh(2\alpha)-
\sin\varphi\sinh(2\chi))(Y_{3}Y_{6}+Y_{4}Y_{5}),
\nonu \\
\widetilde{A}_{1}^{\;27}
& = & 2i(\cosh\chi+e^{i\varphi}\sinh\chi)(\sin\phi\sinh(2\alpha)+
\sin\varphi\sinh(2\chi))(Y_{1}Y_{8}+Y_{2}Y_{7})\nonu \\
 &  & -2i(\cosh\chi-e^{i\varphi}\sinh\chi)(\sin\phi\sinh(2\alpha)-
\sin\varphi\sinh(2\chi))(Y_{3}Y_{6}+Y_{4}Y_{5}),
\nonu \\
\widetilde{A}_{1}^{\;36} 
&= & 2i(\cosh\chi+e^{-i\varphi}\sinh\chi)
(\sin\phi\sinh(2\alpha)-\sin\varphi\sinh(2\chi))(Y_{1}Y_{8}-Y_{2}Y_{7})\nonu
\\
 &  & +2i(\cosh\chi-e^{-i\varphi}\sinh\chi)
(\sin\phi\sinh(2\alpha)+\sin\varphi\sinh(2\chi))(Y_{3}Y_{6}-Y_{4}Y_{5}),
\nonu \\
\widetilde{A}_{1}^{\;45} 
&= & 2i(\cosh\chi+e^{-i\varphi}\sinh\chi)
(\sin\phi\sinh(2\alpha)-\sin\varphi\sinh(2\chi))(Y_{1}Y_{8}-Y_{2}Y_{7})\nonu
\\
 &  &
 -2i(\cosh\chi-e^{-i\varphi}\sinh\chi)
(\sin\phi\sinh(2\alpha)+\sin\varphi\sinh(2\chi))(Y_{3}Y_{6}-Y_{4}Y_{5}).
\label{tilfive}
\eea
Note that using the symmetric property $\widetilde{A}_1^{\;ij} =
\widetilde{A}_1^{\;ji}$, one can obtain the lower triangular parts of
$\widetilde{A}_1$ tensor.

\section{The $28 \times 28$ matrices $u$ and $v$, $A_2$ tensor, 
  kinetic terms, and $SU(8)$ connection 
with $SO(4) \times SO(4)$ symmetry in $SU(8)$ basis}

The 28-beins $u^{IJ}_{\;\;\;KL}$ and $v^{IJKL}$ 
fields 
can be obtained 
by exponentiating the vacuum expectation values $\phi_{ijkl}$ 
(\ref{phiijkl}) via (\ref{calV}). The 28-beins have the
following seven $4 \times 4$ block diagonal matrices $u_i$ and $v_i$
where $i=1, 2, \cdots, 7$ respectively as before.
Each hermitian
submatrix is $4 \times 4$ matrix and we denote
antisymmetric index pairs $[IJ]$ and $[KL]$ explicitly:
\bea
u_{1} & = & \cosh \alpha \; {\bf 1}_{4 \times 4}=
u_{2}=
u_{3},\qquad
u_{4}=
u_{5}=
u_{6}=u_{7} = {\bf 1}_{4 \times 4},
\nonu \\
v_{1} & = & \left(\begin{array}{ccccc}
 & [12] & [34] & [56] & [78]\\{}
[12] & 0 &   e^{-i\zeta}  \sinh \alpha & 0 & 0\\{}
[34] &  & 0 & 0 &  0  \\{}
[56] &  &  & 0 &  e^{i\zeta}  \sinh \alpha \\{}
[78] &  &  &  & 0\end{array}\right)=
-v_{2}= v_{3},
\nonu \\
v_{4} & = &
v_{5}=
v_{6}=
v_{7}=0.
\label{res1}
\eea
See also \cite{CLP} for previous construction of these 28-beins from
the ${\cal N}=4$ gauged $SO(4) =SU(2) \times SU(2)$ truncation of the
full
${\cal N}=8$ gauged $SO(8)$ theory.
The components of $A_2$ tensor, $A_{2, L}^{\;\;\;\;\;IJK}$, 
are obtained similarly
and 
they are classified by two different fields with degeneracies 4, 4 respectively  and  
given by:
\bea
A_{2,\;1}^{\;\;\;234} & = &
A_{2,\;2}^{\;\;\;143}=A_{2,\;3}^{\;\;\;124}=A_{2,\;4}^{\;\;\;132}=  
-e^{-i\zeta}\sinh \alpha =-e^{-i\zeta} \frac{\pa W}{\pa \alpha},
\nonu \\
A_{2,\;5}^{\;\;\;678} & = & 
A_{2,\;6}^{\;\;\;587}=A_{2,\;7}^{\;\;\;568}=A_{2,\;8}^{\;\;\;576}=  
-e^{i\zeta}\sinh \alpha = -e^{i\zeta} \frac{\pa W}{\pa \alpha}.
\label{res2}
\eea
We also rewrite them in terms of the derivative of the superpotential
with respect to $\alpha$.
The kinetic terms 
can be summarized as following seven 
$4\times 4$ block diagonal hermitian matrices 
where each hermitian submatrix can be written as follows:
\bea
A_{\mu,1} & = & \left(\begin{array}{ccccc}
 & [12] & [34] & [56] & [78]\\{}
[12] & 0 & -a^{*} & 0 & 0\\{}
[34] &  & 0 & 0 & 0\\{}
[56] &  &  & 0 & -a\\{}
[78] &  &  &  & 0\end{array}\right)=-
A_{\mu,2}=
A_{\mu,3}, 
\nonu \\
 A_{\mu,4} & = & 0= A_{\mu,5}=A_{\mu,6}=A_{\mu,7}, \qquad
a \equiv \sqrt{2}e^{i\zeta}\left[ \partial_{\mu}\alpha+
\frac{i}{2}\sinh (2\alpha) \partial_{\mu}\zeta \right]. 
\label{res3}
\eea

The quantity in the covariant derivative in (\ref{cov}) is given by
\bea
{\cal B}_{\mu\;\;\;j}^{\;\;i} & = & \mbox{diag}(
-i \partial_{\mu} \zeta \sinh^{2} \alpha,
-i \partial_{\mu} \zeta \sinh^{2} \alpha, 
-i \partial_{\mu} \zeta \sinh^{2} \alpha,
-i \partial_{\mu} \zeta \sinh^{2} \alpha, 
\nonu \\
&& 
i\partial_{\mu}\zeta
\sinh^{2} \alpha,
i\partial_{\mu}\zeta
\sinh^{2} \alpha,
i\partial_{\mu}\zeta
\sinh^{2} \alpha,
i\partial_{\mu}\zeta
\sinh^{2} \alpha).
\label{calB1}
\eea
Note that the above results (\ref{res1}), (\ref{res2}), 
(\ref{res3}), and (\ref{calB1}) 
can be read off from the previous (\ref{uv}), (\ref{A2}), (\ref{amu})
and (\ref{calB}) in appendix A respectively by taking 
$\phi \rightarrow \zeta$ and $\chi \rightarrow 0$.

\section{The $28 \times 28$ matrices $U$ and $V$ and $A_1$ tensor 
with $SO(4) \times SO(4)$ symmetry in $SL(8, {\bf R})$ basis}

Following the notation given in (\ref{abcd}), one obtains
the 28-beins $A_{IJKL}$
\bea
a_2 &= & \left(\begin{array}{ccccc}
 & [13] & [24] & [57] & [68]\\
{}[13] & a & -a & a & -a\\
{}[24] & -b & b & -b & b\\
{}[57] & -a & -a & -a & -a\\
{}[68] & b^{*} & b^{*} & b^{*} & b^{*}\end{array}\right), \qquad
a_4 =\left(\begin{array}{ccccc}
 & [15] & [26] & [37] & [48]\\
{}[15] & -\sqrt{2} & -\sqrt{2} & \sqrt{2} & \sqrt{2}\\
{}[26] & \sqrt{2} & -\sqrt{2} & \sqrt{2} & -\sqrt{2}\\
{}[37] & -\sqrt{2} & \sqrt{2} & \sqrt{2} & -\sqrt{2}\\
{}[48] & \sqrt{2} & \sqrt{2} & \sqrt{2} & \sqrt{2}\end{array}\right),
\nonu \\
a_6 & = & \left(\begin{array}{ccccc}
 & [17] & [28] & [35] & [46]\\
{}[17] & -\sqrt{2} & -\sqrt{2} & -\sqrt{2} & -\sqrt{2}\\
{}[28] & 0 & 0 & 0 & 0\\
{}[35] & \sqrt{2} & -\sqrt{2} & \sqrt{2} & -\sqrt{2}\\
{}[46] & 0 & 0 & 0 & 0\end{array}\right), \qquad
a_1  =  0 = a_3 = a_5 =a_7.
\label{Res1}
\eea
where we introduce 
\bea
a \equiv \sqrt{2}\cosh \alpha,\qquad 
b \equiv \sqrt{2}e^{i\zeta}\sinh \alpha.
\nonu
\eea
Also one has the 28-beins 
$B_{IJ}^{\;\;\;\; KL}$
\bea
b_2 & = & 
\left(\begin{array}{ccccc}
 & [13] & [24] & [57] & [68]\\
{}[13] & a & -a & a & -a\\
{}[24] & b & -b & b & -b\\
{}[57] & -a & -a & -a & -a\\
{}[68] & -b^{*} & -b^{*} & -b^{*} & -b^{*}\end{array}\right),
\qquad
b_4 = \left(\begin{array}{ccccc}
 & [15] & [26] & [37] & [48]\\
{}[15] & -\sqrt{2} & -\sqrt{2} & \sqrt{2} & \sqrt{2}\\
{}[26] & \sqrt{2} & -\sqrt{2} & \sqrt{2} & -\sqrt{2}\\
{}[37] & -\sqrt{2} & \sqrt{2} & \sqrt{2} & -\sqrt{2}\\
{}[48] & \sqrt{2} & \sqrt{2} & \sqrt{2} & \sqrt{2}\end{array}\right),
\nonu \\
b_6 & = & \left(\begin{array}{ccccc}
 & [17] & [28] & [35] & [46]\\
{}[17] & -\sqrt{2} & -\sqrt{2} & -\sqrt{2} & -\sqrt{2}\\
{}[28] & 0 & 0 & 0 & 0\\
{}[35] & \sqrt{2} & -\sqrt{2} & \sqrt{2} & -\sqrt{2}\\
{}[46] & 0 & 0 & 0 & 0\end{array}\right), \qquad
b_1  = 0 = b_3 = b_5 = b_7.
\label{Res2}
\eea
Moreover one has the following results $C_{\;\;\;\;KL}^{IJ}$
\bea
c_2 & = & \left(\begin{array}{ccccc}
 & [13] & [24] & [57] & [68]\\
{}[13] & a & -a & a & -a\\
{}[24] & -b^{*} & b^{*} & -b^{*} & b^{*}\\
{}[57] & -a & -a & -a & -a\\
{}[68] & b & b & b & b\end{array}\right),\qquad
c_4 =\left(\begin{array}{ccccc}
 & [15] & [26] & [37] & [48]\\
{}[15] & -\sqrt{2} & -\sqrt{2} & \sqrt{2} & \sqrt{2}\\
{}[26] & \sqrt{2} & -\sqrt{2} & \sqrt{2} & -\sqrt{2}\\
{}[37] & -\sqrt{2} & \sqrt{2} & \sqrt{2} & -\sqrt{2}\\
{}[48] & \sqrt{2} & \sqrt{2} & \sqrt{2} & \sqrt{2}\end{array}\right),
\nonu \\
c_6 & = & \left(\begin{array}{ccccc}
 & [17] & [28] & [35] & [46]\\
{}[17] & -\sqrt{2} & -\sqrt{2} & -\sqrt{2} & -\sqrt{2}\\
{}[28] & 0 & 0 & 0 & 0\\
{}[35] & \sqrt{2} & -\sqrt{2} & \sqrt{2} & -\sqrt{2}\\
{}[46] & 0 & 0 & 0 & 0\end{array}\right), \qquad
c_1  = 0 = c_3 = c_5 = c_7.
\label{Res3}
\eea
Finally, one gets the 28-beins $D^{IJKL}$
\bea
d_2 & = & \left(\begin{array}{ccccc}
 & [13] & [24] & [57] & [68]\\
{}[13] & -a & a & -a & a\\
{}[24] & -b^{*} & b^{*} & -b^{*} & b^{*}\\
{}[57] & a & a & a & a\\
{}[68] & b & b & b & b\end{array}\right),\qquad
d_4 =\left(\begin{array}{ccccc}
 & [15] & [26] & [37] & [48]\\
{}[15] & \sqrt{2} & \sqrt{2} & -\sqrt{2} & -\sqrt{2}\\
{}[26] & -\sqrt{2} & \sqrt{2} & -\sqrt{2} & \sqrt{2}\\
{}[37] & \sqrt{2} & -\sqrt{2} & -\sqrt{2} & \sqrt{2}\\
{}[48] & -\sqrt{2} & -\sqrt{2} & -\sqrt{2} &
-\sqrt{2}\end{array}\right),\nonu \\ 
d_6 & = & \left(\begin{array}{ccccc}
 & [17] & [28] & [35] & [46]\\
{}[17] & \sqrt{2} & \sqrt{2} & \sqrt{2} & \sqrt{2}\\
{}[28] & 0 & 0 & 0 & 0\\
{}[35] & -\sqrt{2} & \sqrt{2} & -\sqrt{2} & \sqrt{2}\\
{}[46] & 0 & 0 & 0 & 0\end{array}\right), \qquad
d_1  = 0 = d_3 = d_5 = d_7.
\label{Res4}
\eea

The $\widetilde{A}_1$ tensor has the following form:
\bea
\widetilde{A}_{1}^{\;11} & = & 
\widetilde{A}_{1}^{\;22}=\widetilde{A}_{1}^{\;33}=
\widetilde{A}_{1}^{\;44} = 
(\widetilde{A}_{1}^{\;55})^{\ast}  =  
(\widetilde{A}_{1}^{\;66})^{\ast}=
(\widetilde{A}_{1}^{\;77})^{\ast}=
(\widetilde{A}_{1}^{\;88})^{\ast}
\nonu \\
& = & \cosh \alpha +
e^{-i\zeta}\sinh \alpha \left(-Y_{1}^{2}-Y_{2}^{2}-Y_{3}^{2}-
Y_{4}^{2}+Y_{5}^{2}+Y_{6}^{2}+Y_{7}^{2}+Y_{8}^{2}\right),
\nonu \\
\widetilde{A}_{1}^{\;15} & = & 
i\sin \zeta \sinh\left(2\alpha\right)
\left(Y_{1}Y_{5}+Y_{2}Y_{6}-Y_{3}Y_{7}-Y_{4}Y_{8}\right),
\nonu \\
\widetilde{A}_{1}^{\;26} & = & 
i\sin \zeta \sinh\left(2\alpha\right)
\left(Y_{1}Y_{5}+Y_{2}Y_{6}+Y_{3}Y_{7}+Y_{4}Y_{8}\right),
\nonu \\
\widetilde{A}_{1}^{\;37}& = &
i\sin \zeta
\sinh\left(2\alpha\right)\left(-Y_{1}Y_{5}+Y_{2}Y_{6}+Y_{3}Y_{7}-Y_{4}Y_{8}\right),
\nonu \\
\widetilde{A}_{1}^{\;48} & = & 
i\sin \zeta \sinh\left(2\alpha\right)
\left(-Y_{1}Y_{5}+Y_{2}Y_{6}-Y_{3}Y_{7}-Y_{4}Y_{8}\right),
\nonu \\
\widetilde{A}_{1}^{\;16} & = & 
i\sin \zeta \sinh\left(2\alpha\right)
\left(Y_{1}Y_{6}-Y_{2}Y_{5}-Y_{3}Y_{8}+Y_{4}Y_{7}\right),
\nonu \\
\widetilde{A}_{1}^{\;25} & = & 
i\sin \zeta \sinh\left(2\alpha\right)
\left(-Y_{1}Y_{6}+Y_{2}Y_{5}-Y_{3}Y_{8}+Y_{4}Y_{7}\right),
\nonu \\
\widetilde{A}_{1}^{\;38} & = & 
i\sin \zeta \sinh\left(2\alpha\right)
\left(-Y_{1}Y_{6}-Y_{2}Y_{5}+Y_{3}Y_{8}+Y_{4}Y_{7}\right),
\nonu 
\eea
and
\bea
\widetilde{A}_{1}^{\;47}& = &
i\sin \zeta
\sinh\left(2\alpha\right)\left(Y_{1}Y_{6}+Y_{2}Y_{5}+Y_{3}Y_{8}+Y_{4}Y_{7}\right),
\nonu \\
\widetilde{A}_{1}^{\;17} & = & 
i\sin \zeta \sinh\left(2\alpha\right)
\left(Y_{1}Y_{7}-Y_{2}Y_{8}+Y_{3}Y_{5}-Y_{4}Y_{6}\right),
\nonu \\
\widetilde{A}_{1}^{\;28} & = & 
i\sin \zeta \sinh\left(2\alpha\right)
\left(-Y_{1}Y_{7}+Y_{2}Y_{8}+Y_{3}Y_{5}-Y_{4}Y_{6}\right),
\nonu \\
\widetilde{A}_{1}^{\;35} & = & 
i\sin \zeta \sinh\left(2\alpha\right)
\left(Y_{1}Y_{7}+Y_{2}Y_{8}+Y_{3}Y_{5}+Y_{4}Y_{6}\right),
\nonu \\
\widetilde{A}_{1}^{\;46} & = & 
i\sin \zeta \sinh\left(2\alpha\right)
\left(-Y_{1}Y_{7}-Y_{2}Y_{8}+Y_{3}Y_{5}+Y_{4}Y_{6}\right),
\nonu \\
\widetilde{A}_{1}^{\;18} & = & 
i\sin \zeta \sinh\left(2\alpha\right)
\left(Y_{1}Y_{8}+Y_{2}Y_{7}+Y_{3}Y_{6}+Y_{4}Y_{5}\right),
\nonu \\
\widetilde{A}_{1}^{\;27} & = &
i\sin \zeta
\sinh\left(2\alpha\right)\left(Y_{1}Y_{8}+Y_{2}Y_{7}-Y_{3}Y_{6}-Y_{4}Y_{5}\right),
\nonu \\
\widetilde{A}_{1}^{\;36} & = & 
i\sin \zeta \sinh\left(2\alpha\right)
\left(Y_{1}Y_{8}-Y_{2}Y_{7}+Y_{3}Y_{6}-Y_{4}Y_{5}\right),
\nonu \\
\widetilde{A}_{1}^{\;45} & = & 
i\sin \zeta \sinh\left(2\alpha\right)
\left(Y_{1}Y_{8}-Y_{2}Y_{7}-Y_{3}Y_{6}+Y_{4}Y_{5}\right),
\nonu \\
\widetilde{A}_{1}^{\;12} & = & 
\widetilde{A}_{1}^{\;13}=
\widetilde{A}_{1}^{\;14}=\widetilde{A}_{1}^{\;23}=
\widetilde{A}_{1}^{\;24}=
\widetilde{A}_{1}^{\;34}=0.
\label{Res5}
\eea
There is the symmetric property $\widetilde{A}_1^{\;ij} =
\widetilde{A}_1^{\;ji}$. 
Note that the above results (\ref{Res1}), (\ref{Res2}), (\ref{Res3}), (\ref{Res4}) and
(\ref{Res5}) can be read off from the previous (\ref{A}), (\ref{B}),
(\ref{C}), (\ref{D}), (\ref{tilone}), (\ref{tiltwo}),
(\ref{tilthree}), (\ref{tilfour}) and (\ref{tilfive})   
in appendix B by taking 
$\phi \rightarrow \zeta$ and $\chi \rightarrow 0$.



\begin{thebibliography}{99}

\bibitem{ABJM}
  O.~Aharony, O.~Bergman, D.~L.~Jafferis and J.~Maldacena,
  ``N=6 superconformal Chern-Simons-matter theories, M2-branes and their
  gravity duals,''
  JHEP {\bf 0810}, 091 (2008)
  [arXiv:0806.1218 [hep-th]].

\bibitem{AP}
  C.~Ahn and J.~Paeng,
  ``Three-dimensional SCFTs, supersymmetric domain wall and renormalization
  group flow,''
  Nucl.\ Phys.\  B {\bf 595}, 119 (2001)
  [arXiv:hep-th/0008065].

\bibitem{AW}
  C.~Ahn and K.~Woo,
  ``Supersymmetric domain wall and RG flow from 4-dimensional gauged N = 8
  supergravity,''
  Nucl.\ Phys.\  B {\bf 599}, 83 (2001)
  [arXiv:hep-th/0011121].

\bibitem{NW}
  H.~Nicolai and N.~P.~Warner,
  ``The SU(3) X U(1) Invariant Breaking Of Gauged N=8 Supergravity,''
  Nucl.\ Phys.\  B {\bf 259}, 412 (1985).

\bibitem{AI}
  C.~Ahn and T.~Itoh,
  ``An N = 1 supersymmetric G(2)-invariant flow in M-theory,''
  Nucl.\ Phys.\  B {\bf 627}, 45 (2002)
  [arXiv:hep-th/0112010].

\bibitem{AR99}
  C.~Ahn and S.~J.~Rey,
  ``More CFTs and RG flows from deforming M2/M5-brane horizon,''
  Nucl.\ Phys.\  B {\bf 572}, 188 (2000)
  [arXiv:hep-th/9911199].

\bibitem{Warner83}
  N.~P.~Warner,
  ``Some New Extrema Of The Scalar Potential Of Gauged N=8 Supergravity,''
  Phys.\ Lett.\  B {\bf 128}, 169 (1983).

\bibitem{Warner84}
  N.~P.~Warner,
  ``Some Properties Of The Scalar Potential In Gauged Supergravity Theories,''
  Nucl.\ Phys.\  B {\bf 231}, 250 (1984).

\bibitem{AI02}
  C.~Ahn and T.~Itoh,
  ``The 11-dimensional metric for AdS/CFT RG flows with common SU(3)
  invariance,''
  Nucl.\ Phys.\  B {\bf 646}, 257 (2002)
  [arXiv:hep-th/0208137].

\bibitem{CPW}
  R.~Corrado, K.~Pilch and N.~P.~Warner,
  ``An N = 2 supersymmetric membrane flow,''
  Nucl.\ Phys.\  B {\bf 629}, 74 (2002)
  [arXiv:hep-th/0107220].

\bibitem{Ahn0806n2}
  C.~Ahn,
  ``Holographic Supergravity Dual to Three Dimensional N=2 Gauge Theory,''
  JHEP {\bf 0808}, 083 (2008)
  [arXiv:0806.1420 [hep-th]].

\bibitem{BKKS}
  M.~Benna, I.~Klebanov, T.~Klose and M.~Smedback,
  ``Superconformal Chern-Simons Theories and $AdS_4/CFT_3$ Correspondence,''
  JHEP {\bf 0809}, 072 (2008)
  [arXiv:0806.1519 [hep-th]].

\bibitem{KKM}
  I.~Klebanov, T.~Klose and A.~Murugan,
  ``$AdS_4/CFT_3$ -- Squashed, Stretched and Warped,''
  arXiv:0809.3773 [hep-th].

\bibitem{Ahn0806n1}
  C.~Ahn,
  ``Towards Holographic Gravity Dual of N=1 Superconformal Chern-Simons Gauge
  Theory,''
  JHEP {\bf 0807}, 101 (2008)
  [arXiv:0806.4807 [hep-th]].

\bibitem{Ahn0812}
  C.~Ahn,
  ``Comments on Holographic Gravity Dual of N=6 Superconformal Chern-Simons
  Gauge Theory,''
  arXiv:0812.4363 [hep-th].

\bibitem{BHPW}
  N.~Bobev, N.~Halmagyi, K.~Pilch and N.~P.~Warner,
  ``Holographic, N=1 Supersymmetric RG Flows on M2 Branes,''
  arXiv:0901.2736 [hep-th].

\bibitem{GW}
  C.~N.~Gowdigere and N.~P.~Warner,
  ``Flowing with eight supersymmetries in M-theory and F-theory,''
  JHEP {\bf 0312}, 048 (2003)
  [arXiv:hep-th/0212190].

\bibitem{PW}
  C.~N.~Pope and N.~P.~Warner,
  ``A dielectric flow solution with maximal supersymmetry,''
  JHEP {\bf 0404}, 011 (2004)
  [arXiv:hep-th/0304132].

\bibitem{LS}
  R.~G.~Leigh and M.~J.~Strassler,
  ``Exactly Marginal Operators And Duality In Four-Dimensional N=1
  Supersymmetric Gauge Theory,''
  Nucl.\ Phys.\  B {\bf 447}, 95 (1995)
  [arXiv:hep-th/9503121].

\bibitem{FGPW}
  D.~Z.~Freedman, S.~S.~Gubser, K.~Pilch and N.~P.~Warner,
  ``Renormalization group flows from holography supersymmetry and a
  c-theorem,''
  Adv.\ Theor.\ Math.\ Phys.\  {\bf 3}, 363 (1999)
  [arXiv:hep-th/9904017].

\bibitem{BL0711}
  J.~Bagger and N.~Lambert,
  ``Gauge Symmetry and Supersymmetry of Multiple M2-Branes,''
  Phys.\ Rev.\  D {\bf 77}, 065008 (2008)
  [arXiv:0711.0955 [hep-th]].

\bibitem{BL06}
  J.~Bagger and N.~Lambert,
  ``Modeling multiple M2's,''
  Phys.\ Rev.\  D {\bf 75}, 045020 (2007)
  [arXiv:hep-th/0611108].

\bibitem{BL0712}
  J.~Bagger and N.~Lambert,
  ``Comments On Multiple M2-branes,''
  JHEP {\bf 0802}, 105 (2008)
  [arXiv:0712.3738 [hep-th]].

\bibitem{diffsusy}
  D.~Gaiotto and A.~Tomasiello,
  ``The gauge dual of Romans mass,''
  arXiv:0901.0969 [hep-th].

\bibitem{CDS}
  S.~Cherkis, V.~Dotsenko and C.~Saemann,
  ``On Superspace Actions for Multiple M2-Branes, Metric 3-Algebras and their
  Classification,''
  arXiv:0812.3127 [hep-th].
  
\bibitem{BILPSZ}
I.~L.~Buchbinder, E.~A.~Ivanov, O.~Lechtenfeld, N.~G.~Pletnev, 
I.~B.~Samsonov and B.~M.~Zupnik,
  ``ABJM models in N=3 harmonic superspace,''
  arXiv:0811.4774 [hep-th].

\bibitem{Ahn0810}
  C.~Ahn,
  ``Other Squashing Deformation and N=3 Superconformal Chern-Simons Gauge
  Theory,''
  Phys.\ Lett.\  B {\bf 671}, 303 (2009)
  [arXiv:0810.2422 [hep-th]].

\bibitem{ADFGT}
  L.~Andrianopoli, R.~D'Auria, S.~Ferrara, P.~A.~Grassi and M.~Trigiante,
  ``Exceptional N=6 and N=2 $AdS_4$ Supergravity, and Zero-Center Modules,''
  arXiv:0810.1214 [hep-th].

\bibitem{JT}
  D.~L.~Jafferis and A.~Tomasiello,
  ``A simple class of N=3 gauge/gravity duals,''
  JHEP {\bf 0810}, 101 (2008)
  [arXiv:0808.0864 [hep-th]].

\bibitem{CJ}
  E.~Cremmer and B.~Julia,
  ``The N=8 Supergravity Theory. 1. The Lagrangian,''
  Phys.\ Lett.\  B {\bf 80}, 48 (1978).

\bibitem{dN82}
  B.~de Wit and H.~Nicolai,
  ``N=8 Supergravity With Local SO(8) X SU(8) Invariance,''
  Phys.\ Lett.\  B {\bf 108}, 285 (1982).

\bibitem{dN82-1}
  B.~de Wit and H.~Nicolai,
  ``N=8 Supergravity,''
  Nucl.\ Phys.\  B {\bf 208}, 323 (1982).

\bibitem{ST}
  K.~Skenderis and P.~K.~Townsend,
  ``Gravitational stability and renormalization-group flow,''
  Phys.\ Lett.\  B {\bf 468}, 46 (1999)
  [arXiv:hep-th/9909070].

\bibitem{PW1}
  K.~Pilch and N.~P.~Warner,
  ``N = 2 supersymmetric RG flows and the IIB dilaton,''
  Nucl.\ Phys.\ B {\bf 594}, 209 (2001)
  [arXiv:hep-th/0004063].

\bibitem{AW01}
  C.~Ahn and K.~Woo,
  ``Domain wall and membrane flow from other gauged d = 4, n = 8  supergravity.
  I,''
  Nucl.\ Phys.\  B {\bf 634}, 141 (2002)
  [arXiv:hep-th/0109010].

\bibitem{HLL}
  K.~Hosomichi, K.~M.~Lee and S.~Lee,
  ``Mass-Deformed Bagger-Lambert Theory and its BPS Objects,''
  Phys.\ Rev.\  D {\bf 78}, 066015 (2008)
  [arXiv:0804.2519 [hep-th]].

\bibitem{Ahn0809}
  C.~Ahn,
  ``Squashing Gravity Dual of N=6 Superconformal Chern-Simons Gauge Theory,''
  arXiv:0809.3684 [hep-th].

\bibitem{BKW}
  V.~Borokhov, A.~Kapustin and X.~k.~Wu,
  ``Monopole operators and mirror symmetry in three dimensions,''
  JHEP {\bf 0212}, 044 (2002)
  [arXiv:hep-th/0207074].

\bibitem{N4}
  Y.~Imamura and S.~Yokoyama,
  ``N=4 Chern-Simons theories and wrapped M5-branes in their gravity duals,''
  arXiv:0812.1331 [hep-th].

\bibitem{IK}
  Y.~Imamura and K.~Kimura,
  ``N=4 Chern-Simons theories with auxiliary vector multiplets,''
  JHEP {\bf 0810}, 040 (2008)
  [arXiv:0807.2144 [hep-th]].

\bibitem{HLLLP}
  K.~Hosomichi, K.~M.~Lee, S.~Lee, S.~Lee and J.~Park,
  ``N=4 Superconformal Chern-Simons Theories with Hyper and Twisted Hyper
  Multiplets,''
  JHEP {\bf 0807}, 091 (2008)
  [arXiv:0805.3662 [hep-th]].

\bibitem{dWNW}
  B.~de Wit, H.~Nicolai and N.~P.~Warner,
  ``The Embedding Of Gauged N=8 Supergravity Into D = 11 Supergravity,''
  Nucl.\ Phys.\ B {\bf 255}, 29 (1985).

\bibitem{KW}
  A.~Khavaev and N.~P.~Warner,
  ``An N = 1 supersymmetric Coulomb flow in IIB supergravity,''
  Phys.\ Lett.\  B {\bf 522}, 181 (2001)
  [arXiv:hep-th/0106032].

\bibitem{JLP00}
  C.~V.~Johnson, K.~J.~Lovis and D.~C.~Page,
  ``Probing some N = 1 AdS/CFT RG flows,''
  JHEP {\bf 0105}, 036 (2001)
  [arXiv:hep-th/0011166].

\bibitem{JLP01}
  C.~V.~Johnson, K.~J.~Lovis and D.~C.~Page,
  ``The Kaehler structure of supersymmetric holographic RG flows,''
  JHEP {\bf 0110}, 014 (2001)
  [arXiv:hep-th/0107261].

\bibitem{AW02}
  C.~Ahn and K.~Woo,
  ``Domain wall from gauged d = 4, N = 8 supergravity. II,''
  JHEP {\bf 0311}, 014 (2003)
  [arXiv:hep-th/0209128].

\bibitem{ps} 
R. Slansky, Phys. Rep. {\bf 79} (1981) 1.

\bibitem{PS1}
J. Patera and D. Sankoff, Tables of Branching rules for
Representations of Simple Lie Algebras(L'Universit\'{e} de Montr\'{e}al, 
Montr\'{e}al, 1973).

\bibitem{PS2}
W. MacKay and J. Patera, Tables of Dimensions, Indices and 
Branching Rules for representations of Simple Algebras(Dekker, New
York, 1981).

\bibitem{GSP}
  J.~Gomis, A.~J.~Salim and F.~Passerini,
  ``Matrix Theory of Type IIB Plane Wave from Membranes,''
  JHEP {\bf 0808}, 002 (2008)
  [arXiv:0804.2186 [hep-th]].

\bibitem{GRVV}
  J.~Gomis, D.~Rodriguez-Gomez, M.~Van Raamsdonk and H.~Verlinde,
  ``A Massive Study of M2-brane Proposals,''
  JHEP {\bf 0809}, 113 (2008)
  [arXiv:0807.1074 [hep-th]].

\bibitem{CLP}
  M.~Cvetic, H.~Lu and C.~N.~Pope,
  ``Four-dimensional N = 4, SO(4) gauged supergravity from D = 11,''
  Nucl.\ Phys.\  B {\bf 574}, 761 (2000)
  [arXiv:hep-th/9910252].

\bibitem{GNW}
  C.~N.~Gowdigere, D.~Nemeschansky and N.~P.~Warner,
  ``Supersymmetric solutions with fluxes from algebraic Killing spinors,''
  Adv.\ Theor.\ Math.\ Phys.\  {\bf 7}, 787 (2004)
  [arXiv:hep-th/0306097].

\bibitem{BW}
  I.~Bena and N.~P.~Warner,
  ``A harmonic family of dielectric flow solutions with maximal
  supersymmetry,''
  JHEP {\bf 0412}, 021 (2004)
  [arXiv:hep-th/0406145].

\bibitem{LLM}
  H.~Lin, O.~Lunin and J.~M.~Maldacena,
  ``Bubbling AdS space and 1/2 BPS geometries,''
  JHEP {\bf 0410}, 025 (2004)
  [arXiv:hep-th/0409174].

\end{thebibliography}
\end{document}